\documentclass[12pt,letterpaper]{article}
\pdfoutput=1
\usepackage{jheppub}
\usepackage[english]{babel}
\usepackage{amsmath,amssymb,amsbsy,amstext, amsthm}
\usepackage{graphicx}
\usepackage{amsfonts}
\usepackage{amssymb}
\usepackage{subfigure}
\usepackage{float}
\usepackage{stmaryrd}
\usepackage{dsfont}
\usepackage{amssymb,amsmath}
\usepackage{epsfig}
\usepackage{epstopdf}
\usepackage{latexsym}
\usepackage{booktabs}
\usepackage{bbm}

\makeatletter
\def\@fpheader{\relax}
\makeatother

\def\x{{\mathbf{x}}}
\def\R{{\cal R} }
\def\F{{\cal F} }
\def\T{{\cal T} }
\def\O{{\cal O} }
\def\k{{\mathbf{k}}}
\def\p{{\mathbf{p}}}
\def\M{{M_{\text{pl}}}}
\def\({{\left(}}
\def\){{\right)(}}
\def\[{{\left[}}
\def\]{{\right]}}
\def\I{{\cal I} }

\subheader{\begin{flushright}
UTTG-33-13\\
TCC-027-13\\
\end{flushright}}

\title{Non-Gaussianity Consistency Relations, Initial States and Back-reaction}
\author{Sandipan Kundu}
\affiliation{Texas Cosmology Center, University of Texas, Austin, TX 78712}
\affiliation{Theory Group, Department of Physics, University of Texas, Austin, TX 78712}
\emailAdd{sandyk@physics.utexas.edu}

\abstract{ We explore the consistency relations for the three-point functions, in the squeezed limit, of scalar and tensor perturbations in single-field inflation with general initial conditions for the perturbations. For slow-roll inflation, we find that all the three-point functions of scalar and tensor perturbations with a coherent state as the initial state are identical to the three-point functions with the Bunch-Davies initial state. On the other hand, there is an apparent violation of some of the consistency relations for initial states that are related to the Bunch-Davies state by Bogoliubov transformations and we identify the reason for this violation. The back-reaction calculations indicate that the three-point functions for these states can be large enough to violate the consistency relations, however, they are too small to be observed in the near future. }

\begin{document}

\maketitle
\flushbottom


\section{Introduction}

One of the exciting aspects of cosmology is to understand the period of `cosmic inflation' that powered the epoch of the Big Bang. It is truly remarkable that we can address any meaningful question about an epoch shortly after the universe was born.  A period of inflation naturally solves several puzzles of the standard big bang scenario. But the most important success of inflation is that it can explain the temperature fluctuations of cosmic microwave background (CMB) and the large scale structures (LSS) of the universe. 

The physics of inflation is rather simple. A rapid expansion of the universe during inflation is thought to be responsible for a flat and homogeneous universe on the large scale. On the other hand, quantum fluctuations produced during inflation grew via gravitational instability to form structures in the universe. In inflationary cosmological theories the temperature fluctuations of CMB and the large scale structures are directly related to the scalar curvature perturbations produced during inflation.  Inflation naturally  predicts an almost scale invariant power spectrum of primordial fluctuations\cite{Starobinsky:1982ee, Hawking:1982cz, Guth:1982ec, Bardeen:1983qw, Mukhanov:1985rz} which accords with current  observations from the size of the observable universe down to the scales of around a Mpc\cite{Hlozek:2011pc,Ade:2013uln}. Another exciting aspect of inflation is that it also predicts tensor fluctuations that may in the future be directly observed in cosmological gravitational waves\cite{Rubakov:1982df}. 

Despite its great success,  the details of the physics of inflation are still unknown. A large number of models of inflation successfully explain all the  observations making it practically impossible to distinguish between different models. The three-point functions of primordial fluctuations\cite{Maldacena:2002vr, Gao:2012ib, Sreenath:2013xra} are important observables that in principle can be used to differentiate between single-field and multi-field inflation models. In particular,  the three-point functions for single-field inflation, in squeezed limits, obey certain consistency relations that can provide us with an important tool to falsify or establish single-field inflation \cite{Maldacena:2002vr, Creminelli:2004yq,Cheung:2007sv,Creminelli:2011rh,Kundu:2011sg,Senatore:2012wy,Senatore:2012ya,Assassi:2012et,LopezNacir:2012rm,Pimentel:2013gza}.  

Under very general assumptions: (i) it is effectively a single field inflation and (ii) there is no super-horizon evolution of the perturbations (i.e. both scalar and tensor perturbations are frozen outside the horizon), the three-point functions of comoving curvature perturbation $(\R_\k)$ and tensor perturbation $(h^s_{\k})$, in the squeezed limit $k_1, k_2\gg k_3$ are known to obey some consistency relations which are of the form
\begin{align}
&\langle\hat{\R}_{\k_1}\hat{\R}_{\k_2}\hat{\R}_{\k_3}\rangle=(2 \pi)^3 P_{\R}(k_1)P_{\R}(k_3)(n_s-1)\delta^3(\sum\k)\ ,\nonumber\\
&\langle\hat{\R}_{\k_1}\hat{\R}_{\k_2}\hat{h}^s_{\k_3}\rangle=(2 \pi)^3 P_{\R}(k_1)P_{h}(k_3)\left(2-\frac{n_s}{2}\right)\frac{\k_{1;i} \k_{1;j}\epsilon^{s}_{ij}(\k_3)}{ k_1^2}\delta^3(\sum\k)\  ,\nonumber\\
&\langle\hat{h}^s_{\k_1}\hat{h}^{s'}_{\k_2}\hat{\R}_{\k_3}\rangle=(2 \pi)^3 P_{h}(k_1)P_{\R}(k_3)~n_t~\delta_{ss'}\delta^3(\sum\k)\ , \nonumber\\
&\langle\hat{h}^s_{\k_1} \hat{h}^{s'}_{\k_2}\hat{h}^{s''}_{\k_3}\rangle=(2 \pi)^3 P_{h}(k_1)P_{h}(k_3)\delta_{ss'}\left(\frac{3-n_t}{2}\right)\frac{\k_{1;i} \k_{1;j}\epsilon^{s''}_{ij}(\k_3)}{ k_1^2}\delta^3(\sum\k)\ .\nonumber
\end{align}
The other two squeezed limit three-point functions $\langle\hat{h}^s_{\k_1} \hat{\R}_{\k_2}\hat{h}^{s'}_{\k_3}\rangle$, $\langle\hat{h}^s_{\k_1} \hat{\R}_{\k_2}\hat{\R}_{\k_3}\rangle$ vanish in the limit $k_3/k_1\rightarrow 0$.   There has been a great deal of progress in measuring the three-point function (bispectrum) of scalar perturbation from the CMB and  LSS indicating nearly gaussian primordial fluctuations. The current observational constraint on the non-Gaussianity parameter $f_{NL}$\cite{Bartolo:2004if} is very weak and $f_{NL}^{loc}$ remains the best constrained non-Gaussianity parameter: $ f_{NL}^{loc}=2.7\pm 5.8$ (Planck)\cite{Ade:2013ydc}.

From a theoretical point of view, inflation is important because it gives us an opportunity to test predictions of the quantum field theory in a curved space time. One aspect of prime interest  is to understand how much information about the quantum state of primordial fluctuations in the beginning of inflation can be obtained. In order to achieve that  it is essential to explore the dependence of different observables on initial states of the primordial fluctuations. ÊRecently, this area of research has instigated  interest among  physicists mainly because it has the exciting possibility of providing us a window for the physics before inflation. ÊIn this paper, we explore how the single-field consistency relations depend on initial states of the perturbations.  The primary motivation is two-fold: first as the precision of the observations is increasing significantly, we may learn more about the initial state of the fluctuations in the near future. Second, in order for the consistency relations to be applied as a tool to falsify or establish single-field inflation, it is important to know if they are valid for general initial states.

The two-point and three-point functions of primordial fluctuations are generally computed assuming that the fluctuations are initially in the Bunch-Davies state\cite{Bunch:1978yq,Birrell:1982ix}. For slow-roll inflation it is well known that all the consistency relations are obeyed for the Bunch-Davies initial state \cite{Maldacena:2002vr}. There has been a great deal of work focused on departure from the Bunch-Davies state\cite{Kundu:2011sg,Gasperini:1993yf, Holman:2007na, Meerburg:2009ys, Meerburg:2009fi, Das:2009sg, Ashoorioon:2010xg, Agullo:2010ws, Chialva:2011hc, Ganc:2011dy, Carney:2011hz, Giovannini:2012rg, Agarwal:2012mq, Gong:2013yvl, Aravind:2013lra, Flauger:2013hra, Giovannini:2013noa, Ashoorioon:2013eia, Aslanyan:2013jwa, Bahrami:2013isa}. In this paper, we will introduce general initial states of both scalar and tensor perturbations built over the Bunch-Davies state  and consequently compute the two-point and three-point functions of scalar and tensor perturbations for slow-roll inflation. We will show that the consistency relations are obeyed for coherent initial states; in fact all the three-point functions of scalar and tensor perturbations with a coherent state as the initial state are identical to the three-point functions with the Bunch-Davies initial state. It is perhaps not very surprising since one can think of a coherent state as zero-point quantum fluctuations around some classical state. Interactions will generate non-trivial one-point functions, however, that will contribute only to the classical part. The quantum fluctuations contribute to the physically relevant part of the three-point correlations and hence they remain unchanged.

On the other hand, the consistency relation of scalar three-point function is known to be violated for initial states that are related to the Bunch-Davies state by Bogoliubov transformations (we will call them $\alpha$-states)\cite{Ganc:2011dy,Aravind:2013lra,Flauger:2013hra,Bahrami:2013isa}.\footnote{These states are also called Bogoliubov states in the literature because of obvious reason but we think the name $\alpha$-states is infinitesimally better because of their similarity with the $\alpha$-states of de Sitter space.} We will show that when both scalar and tensor perturbations are initially in $\alpha$-states, all four consistency relations are violated. For the derivation of the consistency relations, it is necessary to take the squeezed limit first and then calculate the three-point functions. However, in an honest calculation of the squeezed limit three-point function for a particular model, one should compute the three-point function first and then take the squeezed limit. So, there is an implicit assumption that the terms that are ignored by taking the squeezed limit first are small. However, for $\alpha$-states the correction terms are large in the squeezed limit and hence this assumption is not valid. Let us also note that the other two squeezed limit three point functions $\langle\hat{h}^s_{\k_1} \hat{\R}_{\k_2}\hat{h}^{s'}_{\k_3}\rangle$ and $\langle\hat{h}^s_{\k_1} \hat{\R}_{\k_2}\hat{\R}_{\k_3}\rangle$ remain vanishingly small even for $\alpha$-states because there is still no cross-correlation between scalar and tensor perturbations.

We will then demonstrate that the back-reaction of the initial state of the primordial fluctuations imposes some restrictions on how large the violations can be. The energy density stored in the excited initial state has to be small compare to the kinetic energy of inflation for the slow-roll parameters to be unaffected. In particular, for the scalar three-point function, this imposes a constraint: $f_{NL}^{loc}\lesssim 1$ and hence it is unobservable in the near future \cite{Aravind:2013lra,Flauger:2013hra}; however, it is large enough to violate the consistency relation.

The rest of the paper is organized as follows. In section \ref{conrel} we will introduce non-Gaussianity matrix $\F$ and then present some semiclassical arguments to reproduce the consistency relations. In section \ref{review}, we review quantization of the fluctuations in inflationary universe. Section \ref{genstate} is devoted to the introduction of excited initial states and computation of power spectrums of scalar and tensor perturbations. In section \ref{ng}, we calculate the three-point functions for slow-roll inflation with the Bunch-Davies state and coherent states to demonstrate that the consistency relations are obeyed. On the other hand, in section \ref{alpha}, we show that the consistency relations are violated for $\alpha$-states. We end with a discussion on back-reaction of excited initial state in section \ref{brn} and concluding remarks in section \ref{conclusions}. General results of the three-point functions with $\alpha$-states are relegated to appendix \ref{appendix}.

\section{Consistency relations for the three-point functions}\label{conrel}
In this paper, we will assume that there is effectively one single scalar degree of freedom during inflation; discussion of this section does not particularly depend on the details of the   single-field model. Inflation generates both scalar $(\R)$ and tensor $(h_{ij})$ perturbations and the power spectrums of the perturbations $P_\R$ and $P_h$, defined as\footnote{We will discuss this in more detail in the next section for slow-roll inflation.}
\begin{equation}
\langle\hat{\R}_{\k}\hat{\R}_{\k'}\rangle=(2\pi)^3\delta^3(\k+\k')P_{\R}\ , \qquad
\langle\hat{h}^s_{\k}\hat{h}^{s'}_{\k'}\rangle=(2\pi)^3\delta^3(\k+\k')\delta_{ss'}P_{h} \ ,
\end{equation}
depend on the details of the model. In the last equation, $\R_\k$ and $h^s_\k$ are the Fourier transforms of $\R$ and $h_{ij}$, respectively. Whereas, for single-field inflation the three-point functions of the perturbations, in the squeezed limit (i.e. $k_1, k_2\gg k_3$), are known to obey some consistency relations which are of the form
\begin{equation}\label{cr}
\langle \hat{A}_{\k_1}\hat{B}_{\k_2}\hat{C}_{\k_3}\rangle=(2\pi)^3 \F_{AC}P_A(k_1)P_C(k_3)\delta_{s(A),s(B)}\frac{\epsilon^{s(C)}_{ij}(\k_3)k_{1;i}k_{i;j}}{k_1^2}\delta^3\left(\sum\k\right)\ ,
\end{equation}
where $\hat{A}_{\k},\hat{B}_{\k}, \hat{C}_{\k}$ are either scalar perturbation $\hat{\R}_{\k}$ or tensor perturbation $\hat{h}^s_{\k}$. We have used the notations that for scalar perturbations $s(\R)=0$ and $\epsilon^{0}_{ij}(\k)\equiv\delta_{ij}$. For tensor perturbations, $s(h)$ is the polarization of the mode and $\epsilon^s_{ij}(\k)$ is the polarization tensor that obeys $\epsilon^s_{ii}(\k)=\k^i \epsilon^s_{ij}(\k)=0$ and $\epsilon^s_{ij}(\k)\epsilon^{ s'}_{ij}(\k)=2 \delta_{ss'}$. $\F_{AC}$ is a measure of non-Gaussianity which can be calculated in terms of other observables using some very general arguments.  Note that
\begin{equation}
f_{NL}^{loc}\equiv-\frac{5}{12}\F_{\R\R}\ .
\end{equation}

In this section, we will make some semiclassical arguments to reproduce the consistency relations (\ref{cr}) along with the non-Gaussianity parameters $\F_{AC}$. Our goal is to compute  $\langle \hat{A}_{\k_1}(\tau)\hat{B}_{\k_2}(\tau)\hat{C}_{\k_3}(\tau)\rangle$ in the squeezed limit (i.e. $k_1, k_2\gg k_3$), after $k_1, k_2$ modes have crossed the horizon; thus $k_3$ mode crossed the horizon in the distant past. Let us now clearly state all the assumptions that we are going to make: (1) it is effectively a single field inflation\footnote{This can be easily generalized for single-clock inflations.} and (2) there is no super-horizon evolution of the perturbations and hence both scalar and tensor perturbations are frozen outside the horizon. We will also use the exact squeezed limit $k_3\rightarrow 0$.

Mode $k_3$ crosses the horizon long before modes $k_1$, $k_2$ and hence we can treat mode $k_3$ classically. It will contribute to the background metric and modes $k_1$ and $k_2$ evolve in this perturbed background which is given by
\begin{equation}\label{metriclocal}
ds^2=-dt^2+a^2(t)g^B_{ij}d\x_i d\x_j\ ,
\end{equation}
where, 
\begin{equation}
g^B_{ij}=e^{-2\R_B(\mathbf{x})\delta_{ij}+h_{B;ij}(\mathbf{x})}
\end{equation}
is the contributions from modes far outside the horizon with $\R_B$ and $h_{B;ij}$ are given by
\begin{align}
\R_B(\mathbf{x}, \tau)=& \int_{k\ll k_1, k_2} \frac{d^3\mathbf{k}}{(2\pi)^{3}}  \R_{\mathbf{k}}e^{i \k.\mathbf{x}}\label{rb}\ ,\\
h_{B;ij}(\mathbf{x},\tau)=& \int_{k\ll k_1, k_2} \frac{d^3\mathbf{k}}{(2\pi)^{3}} \sum_{s=+,\times}\epsilon^s_{ij}(\k) h^s_{\mathbf{k}}(\tau)e^{i \k.\mathbf{x}}\ . \label{hb}
\end{align}
Therefore, in the squeezed limit ($k_3\ll k_1\approx k_2$), we can make the approximation 
\begin{equation}
\langle \hat{A}_{\k_1}(\tau)\hat{B}_{\k_2}(\tau)\hat{C}_{\k_3}(\tau)\rangle\approx\langle \langle \hat{A}_{\k_1}(\tau)\hat{B}_{\k_2}(\tau)\rangle_{\k_3}\hat{C}_{\k_3}(\tau)\rangle\ ,
\end{equation}
where, $\langle \hat{A}_{\k_1}(\tau)\hat{B}_{\k_2}(\tau)\rangle_{\k_3}$ is the two-point function in the perturbed background (\ref{metriclocal}). It is clear from the metric (\ref{metriclocal}) that long wavelength mode (neglecting the gradients) is equivalent to a change of coordinates
\begin{equation}\label{rescaling}
\mathbf{x}_i \rightarrow \mathbf{x}'_i=\Lambda_{ij} \mathbf{x}_j\ , \qquad \text{with} \qquad \Lambda_{ij}= e^{-\R_B\delta_{ij}+h_{B;ij}/2}\ .
\end{equation}

\subsection{Three scalars correlator}
First, we need to find out how $\R_{\k}$ transforms under the change of coordinate (\ref{rescaling}).  It is easy to check that Fourier transform of $\R(\x'(\x),\tau)$ is given by
\begin{align}
\int d^3\mathbf{x} ~ \R(\x'(\x),\tau) e^{-i \k.\mathbf{x}}=\det(\Lambda^{-1}) \R_{\Lambda^{-1}\k }\ .
\end{align}
Therefore, under this change of the coordinates (\ref{rescaling}),  $\R_{\k}$ transforms as
\begin{equation}
\R_{\k} \rightarrow \det(\Lambda^{-1}) \R_{\Lambda^{-1}\k }\ , \qquad \text{with} \qquad (\Lambda^{-1})_{ij}=e^{\R_B\delta_{ij}-h_{B;ij}/2}\ .
\end{equation}
Using the identity $\det(e^A)=e^{tr(A)}$ and $tr(h_{ij})=0$, we obtain 
\begin{align}
\det(\Lambda)=&\exp\left[tr(-\R_B\delta_{ij}+h_{B;ij}/2)\right]=\exp(-3 \R_B),\label{rel1}\\
|\Lambda^{-1}\k|=&\left[(\Lambda^{-1})_{ij}(\Lambda^{-1})_{il}\k_j\k_l\right]^{1/2}
\approx k\left(1+ \R_B- h_{B;ij}\frac{\k_i \k_j}{2 k^2}\right).\label{newk}
\end{align}
Also recall that 
\begin{align}
&\delta^3(\Lambda^{-1}\k_1+\Lambda^{-1}\k_2)=\det(\Lambda)\delta^3(\k_1+\k_2)\label{rel2}.
\end{align}
Next we will compute $\langle \hat{\R}_{\k_1}(\tau)\hat{\R}_{\k_2}(\tau)\rangle$ in the perturbed background (\ref{metriclocal}) after  modes $k_1$, $k_2$ cross the horizon. In the unperturbed background, the two point function in the super-horizon limit is given by $\langle\hat{\R}_{\k_1}(\tau)\hat{\R}_{\k_2}(\tau)\rangle \propto \frac{1}{k^{(4-n_s)}}\delta^3(\k_1+\k_2)$, where $n_s$ is the scalar spectral index at $k=k_1$. Now using equations (\ref{rel1}-\ref{rel2}), we finally obtain  
\begin{align}
\langle \hat{\R}_{\k_1}(\tau)&\hat{\R}_{\k_2}(\tau)\rangle_{\k_3}=\det(\Lambda^{-1})^2 \langle \hat{\R}_{\Lambda^{-1}\k_1}(\tau)\hat{\R}_{\Lambda^{-1}\k_2}(\tau)\rangle \nonumber \\
&=\det(\Lambda^{-1})^2(2 \pi)^3 P_{\R}(|\Lambda^{-1}\k_1|)\delta^3(\Lambda^{-1}\k_1+\Lambda^{-1}\k_2) \nonumber \\
&=(2 \pi)^3 P_{\R}(k_1)\frac{\det(\Lambda^{-1})}{\left(1+ \R_B- h_{B;ij}\frac{\k_{1;i} \k_{1;j}}{2 k_1^2}\right)^{4-n_s}}\delta^3(\k_1+\k_2) \nonumber\\
&=(2 \pi)^3 P_{\R}(k_1)\left[1+(n_s-1)\R_B +\left(2-\frac{n_s}{2}\right)h_{B;ij}\frac{\k_{1;i} \k_{1;j}}{ k_1^2}+...\right]\delta^3(\k_1+\k_2).\label{RRprime}
\end{align}
So far we have treated $\R_{\k_3}$ and $h^s_{\k_3}$ as classical fields but now we will promote both $\R_{\k_3}$ and $ h^s_{\k_3}$ to quantum operators.  Replacing $\R_B$ in the last equation  by (\ref{rb}), we obtain
\begin{align}
\langle\hat{\R}_{\k_1}(\tau)&\hat{\R}_{\k_2}(\tau)\hat{\R}_{\k_3}(\tau)\rangle  \approx\langle \langle\hat{\R}_{\k_1}(\tau)\hat{\R}_{\k_2}(\tau)\rangle_{\k_3}\hat{\R}_{\k_3}(\tau)\rangle \nonumber \\
&= -(2 \pi)^3 P_{\R}(k_1)\delta^3(\k_1+\k_2)(1-n_s) \int_{k\ll k_1, k_2} \frac{d^3\mathbf{k}}{(2\pi)^{3}}e^{i \k.\mathbf{x}}\langle \hat{\R}_{\k}(\tau)\hat{\R}_{\k_3}(\tau)\rangle \nonumber \\
& \approx -(2 \pi)^3 P_{\R}(k_1)P_{\R}(k_3)\delta^3(\k_1+\k_2)(1-n_s) \int_{k\ll k_1, k_2}d^3\mathbf{k}e^{i \k.\mathbf{x}}\delta^3(\k+\k_3)\nonumber \\
& \approx (2 \pi)^3 P_{\R}(k_1)P_{\R}(k_3)(n_s-1)\delta^3(\sum\k)\ ,
\end{align}
where, in the squeezed limit $\sum\k\equiv \k_1+\k_2+\k_3\approx \k_1+\k_2$. Comparing the last equation with equation (\ref{cr}), we find $\F_{\R\R}=(n_s-1)$. Therefore, in the squeezed limit we have,
\begin{equation}\label{localfnl}
f_{NL}^{loc}\approx\frac{5}{12}(1-n_s). 
\end{equation}
Note that for running spectral index, $n_s$ in the last equation stands for spectral index at $k=k_1$.

\subsection{Two scalars and a graviton correlator}
We can perform a similar calculation for $\langle\hat{\R}_{\k_1}(\tau)\hat{\R}_{\k_2}(\tau)\hat{h}^s_{\k_3}(\tau)\rangle$. Replacing  $h_{B;ij}$ in equation (\ref{RRprime}) by (\ref{hb}), we can write
\begin{align}
\langle\hat{\R}_{\k_1}(\tau)&\hat{\R}_{\k_2}(\tau)\hat{h}^s_{\k_3}(\tau)\rangle  \approx\langle \langle\hat{\R}_{\k_1}(\tau)\hat{\R}_{\k_2}(\tau)\rangle_{\k_3}\hat{h}^s_{\k_3}(\tau)\rangle \nonumber \\
&=  P_{\R}(k_1)\delta^3(\sum\k)\left(2-\frac{n_s}{2}\right)\frac{\k_{1;i} \k_{1;j}}{ k_1^2}\int_{k\ll k_1, k_2} d^3\mathbf{k}e^{i \k.\mathbf{x}}\sum_{s'=+,\times}\epsilon^{s'}_{ij}(\k)\langle \hat{h}^s_{\k}(\tau)\hat{h}^{s'}_{\k_3}(\tau)\rangle \nonumber \\
& \approx (2 \pi)^3 P_{\R}(k_1)P_{h}(k_3)\delta^3(\sum\k)\left(2-\frac{n_s}{2}\right)\frac{\k_{1;i} \k_{1;j}}{ k_1^2} \int_{k\ll k_1, k_2}d^3\mathbf{k}~ \epsilon^{s}_{ij}(\k)e^{i \k.\mathbf{x}}\delta^3(\k+\k_3)\nonumber \\
& \approx (2 \pi)^3 P_{\R}(k_1)P_{h}(k_3)\left(2-\frac{n_s}{2}\right)\frac{\k_{1;i} \k_{1;j}\epsilon^{s}_{ij}(\k_3)}{ k_1^2}\delta^3(\sum\k)\ .
\end{align}
Comparing the last equation with equation (\ref{cr}), we find 
\begin{equation}
\F_{\R h}=\left(2-\frac{n_s}{2}\right)\ .
\end{equation}
\subsection{Two gravitons and a scalar correlator}
It is straight forward to use a similar argument to compute the other two three-point functions. First let us note that the Fourier transform of $h_{ij}(\x,\tau)$ is given by,
\begin{equation}
\sum_{s=+,\times}\epsilon^s_{ij}(\k) h^s_{\mathbf{k}}(\tau)=\int d^3\mathbf{x} h_{ij}(\x,\tau) ~ e^{-i \k.\mathbf{x}}.
\end{equation}
Under the coordinate transformation (\ref{rescaling}), $h_{ij}(\x,\tau)\rightarrow h_{ij}(\x'(\x),\tau)$ and the fourier transform of $h_{ij}(\x'(\x),\tau)$ is given by,
\begin{equation}
\int d^3\mathbf{x} h_{ij}(\x'(\x),\tau) ~ e^{-i \k.\mathbf{x}}=\det(\Lambda^{-1})\sum_{s'=+,\times}\epsilon^{s'}_{ij}(\Lambda^{-1}\k) h^{s'}_{\Lambda^{-1}\mathbf{k}}(\tau).
\end{equation}
Therefore, under this coordiante transformation, $ h^s_{\mathbf{k}}(\tau)$ transforms as
\begin{align}
 h^s_{\mathbf{k}}(\tau)\rightarrow \frac{1}{2} \det(\Lambda^{-1})\sum_{s'=+,\times}\epsilon^{s'}_{ij}(\Lambda^{-1}\k)\epsilon^{s}_{ij}(\k) h^{s'}_{\Lambda^{-1}\mathbf{k}}(\tau).
\end{align}
In the unperturbed background, the two point function of gravitons in the super-horizon limit can be written as $\langle\hat{h}^s_{\k_1}(\tau) \hat{h}^{s'}_{\k_2}(\tau)\rangle \propto \frac{1}{k^{(3-n_t)}}\delta^3(\k_1+\k_2)$, where $n_t$ is the tensor spectral index. But we need to compute $\langle\hat{h}^s_{\k_1}(\tau) \hat{h}^{s'}_{\k_2}(\tau)\rangle$ in the perturbed background (\ref{metriclocal}) after modes $k_1$, $k_2$ cross the horizon. Now, using the last equation, we obtain
\begin{align}
\langle\hat{h}^s_{\k_1}&(\tau) \hat{h}^{s'}_{\k_2}(\tau)\rangle_{\k_3}\nonumber\\
=&\frac{1}{4}\det(\Lambda^{-1})^2 \sum_{s_1,s_2=+,\times}\epsilon^{s_1}_{ij}(\Lambda^{-1}\k_1)\epsilon^{s}_{ij}(\k_1)\epsilon^{s_2}_{kl} (\Lambda^{-1}\k_2)\epsilon^{s'}_{kl}(\k_2)\langle \hat{h}^{s_1}_{\Lambda^{-1}\k_1} (\tau)\hat{h}^{s_2}_{\Lambda^{-1}\k_2}(\tau)\rangle \nonumber \\
=&\frac{(2 \pi)^3}{4}\det(\Lambda^{-1}) P_{h}(|\Lambda^{-1}\k_1|)\epsilon^{s}_{ij}(\k_1)\epsilon^{s'}_{kl}(\k_1)\sum_{s_1=+,\times}\epsilon^{s_1}_{ij}(\Lambda^{-1}\k_1)\epsilon^{s_1}_{kl} (\Lambda^{-1}\k_1)\delta^3(\k_1+\k_2)\nonumber\\
=&\frac{(2 \pi)^3}{4} P_{h}(k_1)\epsilon^{s}_{ij}(\k_1)\epsilon^{s'}_{kl}(\k_1)\Pi_{ij,kl}(\Lambda^{-1}\k_1)\frac{\det(\Lambda^{-1})}{\left(1+ \R_B- h_{B;ij}\frac{\k_{1;i} \k_{1;j}}{2 k_1^2}\right)^{3-n_t}}\delta^3(\k_1+\k_2) \nonumber\\
=& \frac{(2 \pi)^3}{4} P_{h}(k_1)\epsilon^{s}_{ij}(\k_1)\epsilon^{s'}_{kl}(\k_1)\Pi_{ij,kl}(\Lambda^{-1}\k_1)\nonumber\\
&~~~~~~~~~~~~~~~~~~~~\times\left[1+n_t\R_B +\left(\frac{3}{2}-\frac{n_t}{2}\right)h_{B;ij}\frac{\k_{1;i} \k_{1;j}}{ k_1^2}+...\right]\delta^3(\k_1+\k_2)\ ,\label{hh1}
\end{align}
 where, $\Pi_{ij,lm}(\k)$ is defined as
\begin{equation}
\Pi_{ij,lm}(\k)=\sum_{s=+,\times}\epsilon^{s}_{ij}(\k)\epsilon^{s}_{lm} (\k).
\end{equation}
A formula can be obtained for $\Pi_{ij,lm}(\k)$ by using the conditions that $\Pi_{ij,lm}(\k)$ is a tensor function of $\hat{\k}$ (because polarization tensor $\epsilon^{s}_{ij}(\k)$ depends only on the direction of vector $\k$), symmetric in $i$ and $j$ and in $l$ and $m$ and $\Pi_{ij,lm}(\k)=\Pi_{lm,ij}(\k)$. $\Pi_{ij,lm}(\k)$ also obeys the conditions $\k_i\Pi_{ij,lm}(\k)=0$ and $\Pi_{ij,ij}(\k)=4$. The last condition comes from the normalization of the polarization tensor $\epsilon^{s}_{ij}(\k)$. Finally we have,
\begin{align}
\Pi_{ij,lm}(\k)=&\delta_{il}\delta_{jm}+\delta_{im}\delta_{jl}-\delta_{ij}\delta_{lm}+ \delta_{ij}\hat{\k}_l\hat{\k}_m + \delta_{lm}\hat{\k}_i\hat{\k}_j\nonumber\\
&-\delta_{il}\hat{\k}_j\hat{\k}_m- \delta_{im}\hat{\k}_j\hat{\k}_l-\delta_{jl}\hat{\k}_i\hat{\k}_m -\delta_{jm}\hat{\k}_i\hat{\k}_l + \hat{\k}_i\hat{\k}_j \hat{\k}_l\hat{\k}_m.
\end{align}
That leads to
\begin{equation}
\epsilon^{s}_{ij}(\k_1)\epsilon^{s'}_{kl}(\k_1)\Pi_{ij,kl}(\Lambda^{-1}\k_1)=4 \delta_{ss'}+ O(h_{ij}^2)
\end{equation}
yielding
\begin{equation}\label{hhprime}
\langle\hat{h}^s_{\k_1}(\tau) \hat{h}^{s'}_{\k_2}(\tau)\rangle_{\k_3}=(2 \pi)^3 P_{h}(k_1)\delta_{ss'} \left[1+n_t\R_B +\left(\frac{3}{2}-\frac{n_t}{2}\right)h_{B;ij}\frac{\k_{1;i} \k_{1;j}}{ k_1^2}+...\right]\delta^3(\k_1+\k_2).
\end{equation}
Again, we will promote both $\R_{\k_3}$ and $ h^s_{\k_3}$ to quantum operators. Using equations (\ref{hb}) and (\ref{hhprime}), we obtain
\begin{align}
\langle\hat{h}^s_{\k_1}(\tau) \hat{h}^{s'}_{\k_2}(\tau)\hat{\R}_{\k_3}(\tau)\rangle&  \approx\langle \langle\hat{h}^s_{\k_1}(\tau) \hat{h}^{s'}_{\k_2}(\tau)\rangle_{\k_3}\hat{\R}_{\k_3}(\tau)\rangle \nonumber \\
&=(2 \pi)^3 P_{h}(k_1)\delta_{ss'}\delta^3(\sum\k)n_t \int_{k\ll k_1, k_2} \frac{d^3\mathbf{k}}{(2\pi)^{3}}e^{i \k.\mathbf{x}}\langle \hat{\R}_{\k}(\tau)\hat{\R}_{\k_3}(\tau)\rangle \nonumber \\
& \approx (2 \pi)^3 P_{h}(k_1)P_{\R}(k_3)\delta_{ss'}\delta^3(\sum\k)n_t \int_{k\ll k_1, k_2}d^3\mathbf{k}e^{i \k.\mathbf{x}}\delta^3(\k+\k_3)\nonumber \\
& \approx (2 \pi)^3 P_{h}(k_1)P_{\R}(k_3)~n_t~\delta_{ss'}\delta^3(\sum\k)\ ,
\end{align}
where, in the squeezed limit $\sum\k\equiv \k_1+\k_2+\k_3\approx \k_1+\k_2$. Comparing the last equation with equation (\ref{cr}), we find $\F_{h\R}=n_t$. Note that for running spectral index, $n_t$ in the last equation stands for the tensor spectral index at $k = k_1$.
\subsection{Three gravitons correlator}
Using equations (\ref{hb}) and (\ref{hhprime}), we obtain
\begin{align}
\langle\hat{h}^s_{\k_1}(\tau) \hat{h}^{s'}_{\k_2}(\tau)\hat{h}^{s''}_{\k_3}(\tau)\rangle  &\approx\langle \langle\hat{h}^s_{\k_1}(\tau) \hat{h}^{s'}_{\k_2}(\tau)\rangle_{\k_3}\hat{h}^{s''}_{\k_3}(\tau)\rangle \nonumber \\
&= (2 \pi)^3 P_{h}(k_1)\delta_{ss'}\delta^3(\sum\k)\left(\frac{3-n_t}{2}\right)\frac{\k_{1;i} \k_{1;j}}{ k_1^2}\nonumber\\
&~~~~~~~~~~~~~~~\times\int_{k\ll k_1, k_2} \frac{d^3\mathbf{k}}{(2\pi)^{3}}e^{i \k.\mathbf{x}}\sum_{s_1=+,\times}\epsilon^{s_1}_{ij}(\k)\langle \hat{h}^{s''}_{\k}(\tau)\hat{h}^{s_1}_{\k_3}(\tau)\rangle \nonumber \\
& \approx (2 \pi)^3 P_{h}(k_1)P_{h}(k_3)\delta_{ss'}\left(\frac{3-n_t}{2}\right)\frac{\k_{1;i} \k_{1;j}\epsilon^{s''}_{ij}(\k_3)}{ k_1^2}\delta^3(\sum\k).
\end{align}
Comparing the last equation with equation (\ref{cr}), we find 
\begin{equation}
\F_{h h}=\left(\frac{3-n_t}{2}\right)\ ,
\end{equation}
with $n_t$ being the tensor spectral index at $k=k_1$.

\

From the discussion of this section it is also obvious that both $\langle\hat{h}^s_{\k_1}(\tau) \hat{\R}_{\k_2}(\tau)\hat{h}^{s'}_{\k_3}(\tau)\rangle$ and $\langle\hat{h}^s_{\k_1}(\tau) \hat{\R}_{\k_2}(\tau)\hat{\R}_{\k_3}(\tau)\rangle$ are zero\footnote{We mean 
\begin{equation}
\frac{\langle\hat{h}^s_{\k_1}(\tau) \hat{\R}_{\k_2}(\tau)\hat{h}^{s'}_{\k_3}(\tau)\rangle}{P_\R(k_2)P_h(k_3)}\ \& \ \frac{\langle\hat{h}^s_{\k_1}(\tau) \hat{\R}_{\k_2}(\tau)\hat{\R}_{\k_3}(\tau)\rangle}{P_\R(k_3)P_h(k_1)}\rightarrow  0
\end{equation}
in the limit $k_3/k_1\rightarrow 0$.} in the squeezed limit $k_1, k_2\gg k_3$ because there is no cross-correlation between scalar and tensor perturbations, i.e.,  $\langle\hat{h}^s_{\k_1}(\tau) \hat{\R}_{\k_2}(\tau)\rangle=0$. Therefore, three-point functions of scalar and tensor perturbations in the squeezed limit, indeed obey consistency relations of the form (\ref{cr}) where $\F_{AC}$ is given 
\begin{equation}\label{F}
\F\equiv
 \begin{pmatrix}
  \F_{\R\R}&~~~~~  \F_{\R h}  \\
  \F_{h\R} &~~~~ \F_{hh} 
 \end{pmatrix}
 =
  \begin{pmatrix}
  n_s-1 ~~~& 2-\frac{n_s}{2}  \\
  n_t & \frac{3-n_t}{2}
 \end{pmatrix}\ ,
\end{equation}
where all the quantities are evaluated at $k=k_1$.

It is important to note that there is an implicit assumption in the derivation of the consistency relations which plays a crucial role. In our derivation, we have taken the squeezed limit $k_3\rightarrow 0$ first and that allows us to approximate the effect of $k_3$-mode as a perturbation to the background metric (\ref{metriclocal}). But in an honest calculation of squeezed limit three-point function, one should compute the three-point function first and then take the squeezed limit $k_3\rightarrow 0$. So, we have made the assumption that the terms that we ignored by taking the squeezed limit first are small. However, we will show that this assumption is not always valid when  the perturbations are in excited initial states.
\section{Quantization of the fluctuations in inflationary universe}\label{review}

\subsection{Scalar field in FRW universe}
Before we proceed to check the validity of the consistency relations for slow-roll inflation with different initial states, let us first review the quantization of fluctuations in inflationary universe. We start with the Lagrangian of gravity and a minimally coupled real scalar field with a canonical kinetic term
\begin{equation}\label{EHaction}
 S=\frac{1}{2}\int \sqrt{-g}d^4x\left[\frac{1}{8\pi G}R - g^{\mu\nu}\partial_\mu \phi \partial_\nu \phi -2V(\phi)\right].
\end{equation}
A homogeneous background solution has the form
\begin{equation}
 ds^2= -dt^2+ a^2(t)d\mathbf{x}^2
\end{equation}
with a background scalar field $\phi(\mathbf{x}, t)= \bar{\phi}(t)$. This background obeys the equations
\begin{align}
& 3H^2= 8 \pi G\left[\frac{1}{2}\dot{\bar{\phi}}^2 +V(\bar{\phi})\right],\\
&\ddot{\bar{\phi}}+3H\dot{\bar{\phi}}+V'(\bar{\phi})=0,
\end{align}
where $H$ is the Hubble parameter $H=\dot{a}/a$. For slow-roll inflation $V(\bar{\phi})$ is approximately constant and slow roll parameters $|\epsilon|,|\eta |\ll 1$, where
\begin{align}\label{slowroll}
\epsilon=&-\frac{\dot{H}}{H^2}\approx\frac{\dot{\bar{\phi}}^2}{2H^2\M^2}=\frac{1}{16 \pi G}\left(\frac{V'}{V}\right)^2 ,\\
\eta=&\frac{1}{8 \pi G}\frac{V''}{V}.
\end{align}
Next we consider perturbations around the homogeneous background solutions
\begin{equation}
  \phi(\mathbf{x}, t)= \bar{\phi}(t)+ \delta \phi(\mathbf{x}, t)
\end{equation}
and the perturbed metric with scalar and tensor perturbations is given by
\begin{equation}
 ds^2=-(1+2\Phi)dt^2+2a(t)(\partial_{i}B) dx^idt+a^2(t)[(1-2\Psi)\delta_{ij}+2\partial_{ij}E + h_{ij}]dx^idx^j.
\end{equation}
Where $h_{ij}$ is purely tensor perturbation and satisfies following conditions:
\begin{equation}
h_{ij}=h_{ji}, \qquad h_{ii}=0, \qquad \partial_i h_{ij}=0.
\end{equation}
Tensor perturbations are gauge-invariant at linear order but scalar perturbations are not. We can avoid fictitious gauge modes of scalar perturbation by introducing gauge-invariant variables\cite{Bardeen:1980kt, Lyth:1984gv}. One such variable is the comoving curvature perturbation 
\begin{equation}\label{curvpert}
\R= \Psi+\frac{H}{\dot{\bar{\phi}}}\delta\phi.
\end{equation}
 Expanding the action (\ref{EHaction}), we get the gauge-invariant second order actions for scalar and tensor perturbations (with conformal time $\tau$ defined in the usual way)
\begin{align}\label{secondaction}
S_{2}^{(s)}&=\frac{1}{2} \int d\tau d^3x a^2 \frac{\dot{\bar{\phi}}^2}{H^2}\left[{\R'}^2-(\partial_i \R)^2 \right]\ ,\\
S_{2}^{(t)}&=\frac{\M^2}{8} \int d\tau d^3x a^2 \left[{h'}_{ij}^2-(\partial_l h_{ij})^2 \right]\ .
\end{align}
Where, Planck mass $\M=(8 \pi G)^{-1/2}$ and $(...)'=\partial_\tau(...)$. We can define the Fourier transforms of the fields in the standard way,
\begin{align}
\R(\mathbf{x}, \tau)=& \int \frac{d^3\mathbf{k}}{(2\pi)^{3}}  \R_{\mathbf{k}}(\tau)e^{i \k.\mathbf{x}}\ ,\\
h_{ij}(\mathbf{x}, \tau)=&\int \frac{d^3\mathbf{k}}{(2\pi)^{3}}  \sum_{s=+,\times}\epsilon^s_{ij}(\k) h^s_{\mathbf{k}}(\tau)e^{i \k.\mathbf{x}} ,
\end{align}
where, $\epsilon^s_{ij}(\k)$ is a real tensor (polarization tensor)\footnote{Note that $\epsilon^s_{ij}(\k)$ depends only on the unit vector $\hat{\k}$.} and it obeys $\epsilon^s_{ii}(\k)=\k^i \epsilon^s_{ij}(\k)=0$ and $\epsilon^s_{ij}(\k)\epsilon^{ s'}_{ij}(\k)=2 \delta_{ss'}$. Because the fields $\R(\mathbf{x}, \tau)$ and $h_{ij}(\mathbf{x}, \tau)$ are real, we have the conditions:
\begin{equation}
\R^*_{\k}(\tau)=\R_{-\k}(\tau)\ , \qquad h^{* s}_{\mathbf{k}}(\tau)=h^{ s}_{\mathbf{-k}}(\tau)\ , \qquad \epsilon^s_{ij}(\k)=\epsilon^s_{ij}(-\k)\ .
\end{equation}
In terms of canonically normalized fields
\begin{equation}
v^0_{\k}(\tau)\equiv \frac{a(\tau)\dot{\bar{\phi}}}{H} \R_{\mathbf{k}}(\tau) \ , \qquad v^s_{\k}(\tau)\equiv \frac{a(\tau)}{\sqrt{2}}\M h^s_{\mathbf{k}}(\tau)
\end{equation}
the action $S_2\equiv S_2^{(s)}+S_2^{(t)}$ becomes
\begin{equation}
S_2=\sum_{s=0,+,\times}\frac{1}{2}\frac{1}{(2 \pi)^3}\int d\tau d^3\k \left[{v'}^s_{\k}(\tau){v_{\k}^{s *}}'(\tau)-k^2 v^s_{\k}(\tau)v^{s*}_{\k}(\tau)+\frac{a''}{a}v^s_{\k}(\tau)v^{s*}_{\k}(\tau)\right]\ . 
\end{equation}
Note that the sum in the last equation is over $s=0,+,\times$, where $s=0$ corresponds to the scalar perturbations and $s=+,\times$ correspond to two polarization modes of the tensor perturbations. From this action we get the following equation for $v^s_{\k}$
\begin{equation}\label{mukhanovequation}
{v^s}_{\k}''(\tau)+ \omega_k^2(\tau) v^s_{\k}(\tau)=0  \qquad s=0,+,\times
\end{equation}
with $ \omega_k^2(\tau)= k^2-(a''/a)$. Let $u_k(\tau)$ and $u_k^*(\tau)$ be linearly independent  complex solutions of equations of motion (\ref{mukhanovequation}). Wronskian $W[u_k,u_k^*]=2 i Im[u'_k(\tau) u^*_k(\tau)]\neq 0$ and it is time-independent; so we can always normalize the mode function $u_k(\tau)$ by the condition 
\begin{equation}\label{normalization}
Im[u'_k(\tau) u^*_k(\tau)]=1. 
\end{equation}
The general solution of equation (\ref{mukhanovequation}) can be written as
\begin{equation}\label{generalsolution}
v^s_{\k}(\tau)=\frac{1}{\sqrt{2}}\left[a^{s-}_{\k}u^*_k(\tau)+ a^{s+}_{-\k}u_k(\tau)\right],
\end{equation}
where $a^{s-}_{\k}$ and $a^{s+}_{-\k}$ are independent of $\tau$ and $ a^{s+}_{\k}= (a^{s-}_{\k})^*$.
\subsection{Quantization of fluctuations}
We will work in the Heisenberg picture to quantize fields $v^s(\mathbf{x}, \tau)$, (where $s=0,+,\times$). We introduce the commutation relations
\begin{equation}\label{commutation1}
\left[\hat{v}^s(\mathbf{x}, \tau), \hat{\pi}^{s'}(\mathbf{y}, \tau)\right]=i \delta^3(\mathbf{x}-\mathbf{y})\delta_{s s'}, 
\end{equation}
where $\hat{\pi}^s= \hat{v}^s{'}$ is the canonical momentum. Now the equation (\ref{generalsolution}) becomes
\begin{equation}\label{quantumv}
 \hat{v}^s_{\k}(\tau)=\frac{1}{\sqrt{2}}\left[{\hat{a}^s}_{\k}u^*_k(\tau)+ \hat{a}^{s\dagger}_{-\k}u_k(\tau)\right] \qquad \text{for} \qquad s=0,+,\times.
\end{equation}
The commutation relations (\ref{commutation1}) lead to commutation relations between $\hat{a}^{s\dagger}_{\k}$ and ${\hat{a}^s}_{\k}$ 
\begin{equation}\label{commutation}
\left[{\hat{a}^s}_{\k_1},\hat{a}^{s'\dagger}_{\k_2}\right]=(2 \pi)^3 \delta^3(\k_1-\k_2)\delta_{s s'},  \qquad   \left[\hat{a}^{s'\dagger}_{\k_1},\hat{a}^{s\dagger}_{\k_2}\right]=\left[{\hat{a}^{s'}}_{\k_1},{\hat{a}^s}_{\k_2}\right]=0 .
\end{equation}
The Hamiltonian of the system is
\begin{equation}\label{hamiltonian}
\hat{H}(\tau)=\frac{1}{4}\frac{1}{(2 \pi)^3}\sum_{s=0,+,\times}\int d^3\k \left[{\hat{a}^s}_{\k}{\hat{a}^s}_{-\k} F_k^*(\tau) + \hat{a}^{s\dagger}_{\k}\hat{a}^{s\dagger}_{-\k} F_k(\tau) + \left({\hat{a}^s}_{\k}\hat{a}^{s\dagger}_{\k}+\hat{a}^{s\dagger}_{\k}{\hat{a}^s}_{\k}\right) E_k(\tau)\right],
\end{equation}
where,
\begin{equation}
F_k(\tau)= (u_k')^2+ \omega_k^2 u_k^2,  \qquad   E_k(\tau)= |u_k'|^2+ \omega_k^2 |u_k^2|.
\end{equation}
Note that the sum in (\ref{hamiltonian}) is over $s=0,+,\times$ and hence it contains both scalar and tensor fluctuations. 
\subsection{Bunch-Davies vacuum}
Next we will define a ``vacuum" state and find out the mode-function that describe the state. The Hamiltonian explicitly depends on the conformal time $\tau$, making it impossible to define a vacuum in a time-independent way. We can define a vacuum by the standard condition: for all $\k$ 
\begin{equation}\label{vaccond1}
{\hat{a}^s}_{\k}|0\rangle=0 \qquad \text{for} \qquad s=0,+,\times\ .
\end{equation}
But this is not sufficient to specify the mode-function. There is no time-independent eigenstate of the Hamiltonian, so we take a particular moment $\tau=\tau_0$, and define vacuum as the lowest-energy eigenstate of the instantaneous Hamiltonian of the fluctuations at $\tau=\tau_0$ (we can always do that as long as $\omega_k^2(\tau_0)\geq0$). That gives us the following initial conditions for the mode function 
\begin{equation}\label{initialcondition}
u_k'(\tau_0)=\pm i \sqrt{\omega_k(\tau_0)}e^{i \lambda(k)},  \qquad  u_k(\tau_0)=\pm  \frac{1}{\sqrt{\omega_k(\tau_0)}}e^{i \lambda(k)},
\end{equation}
where $\lambda(k)$ is some arbitrary time independent function of $k$. In the limit when $\tau_0$ represents infinite past (i.e. $\tau_0\rightarrow -\infty$), this vacuum is called the Bunch-Davies vacuum state. In this limit, $\omega_k^2=k^2\geq0$ and we can define vacuum by equation(\ref{vaccond1}) for all modes.

\subsection{Power-spectrum for slow-roll inflation}
For slow-roll inflation, $V(\bar{\phi})$ is approximately constant and the slow roll parameters $|\epsilon|,|\eta |\ll 1$. Therefore, the equation of state parameter $w\approx-1$ and $(a''/a)= (2/\tau^2)$.\footnote{See \cite{Ackerman:2010he} for constraints on $w$.} Solving equation (\ref{mukhanovequation}) with normalization condition (\ref{normalization}) and initial conditions (\ref{initialcondition}) (with the $+$ sign and $\tau_0\rightarrow -\infty$ limit), we get 
\begin{equation}\label{modesBD}
u_k(\tau)= \frac{e^{ik\tau}}{\sqrt{k}}\left(1+\frac{i}{k\tau}\right). 
\end{equation}
With this mode function we can now compute power-spectrums of scalar and tensor perturbations for slow-roll inflation. 

\subsubsection{Scalar power-spectrum}
Let us first compute the following quantity
\begin{equation}\label{vk2}
\langle\hat{v}^s_{\k}(\tau)\hat{v}^{s'}_{\k'}(\tau)\rangle\equiv\langle 0|\hat{v}^s_{\k}(\tau)\hat{v}^{s'}_{\k'}(\tau)|0\rangle=\frac{1}{2} (2 \pi)^3 \delta^3(\k+\k')\delta_{ss'}|u_k(\tau)|^2, 
\end{equation}
where we got the last equation using (\ref{quantumv}). Before we proceed let us introduce some standard quantities
\begin{equation}\label{ns}
\langle\hat{\R}_{\k}(\tau)\hat{\R}_{\k'}(\tau)\rangle=(2\pi)^3\delta^3(\k+\k')P_{\R},  \qquad  \Delta^2_{\R}=\frac{k^3}{2\pi^2}P_{\R}, \qquad n_s-1=\frac{d\ln \Delta^2_{\R}}{d \ln k} ,
\end{equation}
where $n_s$ is called the scalar spectral index or tilt. Using equations (\ref{modesBD}-\ref{ns}), in the superhorizon limit ($|k\tau|\ll1$), we obtain \cite{Maldacena:2002vr}
\begin{equation}
    \Delta^2_{\R}=\frac{ H^4}{4\pi^2~ \dot{\bar{\phi}}^2 }\ , \qquad  n_s=1-6\epsilon +2\eta\ .
\end{equation}
Where $H$ is the Hubble parameter during inflation. Therefore slow-roll inflation predicts  an almost scale-invariant  scalar power spectrum (i.e. $n_s\approx 1$) which agrees with observation of the CMB and LSS.

\subsubsection{Tensor power-spectrum}
The power spectrum of two polarizations of $h_{ij}$ is defined as
\begin{equation}\label{nt}
\langle\hat{h}^s_{\k}(\tau)\hat{h}^{s'}_{\k'}(\tau)\rangle=(2\pi)^3\delta^3(\k+\k')\delta_{ss'}P_{h},  \qquad  \Delta^2_{h}=\frac{k^3}{2\pi^2}P_{h}.
\end{equation}
The power spectrum for tensor perturbations is defined as the sum of the power spectrum for the two polarizations
\begin{equation}
\Delta^2_{t}\equiv 2\Delta^2_{h}.
\end{equation}
Spectral index  $n_t$ for the tensor perturbations is defined in the following way
\begin{equation}
 n_t=\frac{d\ln \Delta^2_{t}}{d \ln k}.
\end{equation}
Finally, in the superhorizon limit ($|k\tau|\ll1$), we get \cite{Maldacena:2002vr}
\begin{equation}
  \Delta^2_{t}=\frac{ H^2}{\pi^2~ \M^2 }\ ,   \qquad  n_t=-2\epsilon\ .
\end{equation}
Therefore, slow-roll inflation also predicts nearly scale-invariant tensor power-spectrum. Amplitude of tensor power-spectrum is rather small and the tensor-to-scalar ratio is given by
\begin{equation}
r\equiv \frac{\langle \hat{h}_{\ij}(\x)\hat{h}_{\ij}(\x) \rangle}{\langle \hat\R(\x)\hat\R(\x)\rangle}=\frac{4\Delta^2_{h}}{\Delta^2_{\R}}=16\epsilon\ .
\end{equation}
Current bound on tensor-to-scalar ratio is $r<0.12$ (Planck+WP)\cite{Ade:2013uln}. Detection of primordial gravitational waves will provide an important test for slow-roll inflation because there is a consistency relation between $n_t$ and $r$
\begin{equation}
r=-8n_t\ .
\end{equation}


\section{General initial states}\label{genstate}
As a next step, following the discussion of \cite{Kundu:2011sg}, we will define general initial states built over the Bunch-Davies vacuum state $|0\rangle$. It is important to note that we are in the Heisenberg picture where states are time-independent. We can use $\hat{a}^{s\dagger}_\k$ operators to build excited states over the Bunch-Davies vacuum state  
\begin{equation}\label{excitedstate}
|\psi_s\rangle= \frac{1}{\sqrt{n_1!n_2!...}} \left[\left(\hat{a}^{s\dagger}_{\k_1}\right)^{n_1}\left(\hat{a}^{s \dagger}_{\k_2}\right)^{n_2}...\right]|0\rangle. 
\end{equation}
Again note that we have used index $s$ to denote both scalar and tensor perturbations; $s=0$ corresponds to scalar perturbations and $s=\times,+$ correspond to two polarization modes of tenser perturbations. We can write down a general excited state for a perturbation, using equation (\ref{excitedstate})
\begin{equation}\label{generalstate}
|G_s\rangle=\sum_{\psi_s}C_{\psi_s}|\psi_s\rangle. 
\end{equation}
Therefore, a general state can be written as a direct product 
\begin{equation}\label{Gstate}
|G\rangle=|G_{s=0}\rangle\varotimes |G_{s=+}\rangle\varotimes |G_{s=\times}\rangle.
\end{equation}
And with this initial state we can compute the power spectrum using
\begin{equation}
\langle\hat{v}^s_{\k}(\tau)\hat{v}^{s'}_{\k'}(\tau)\rangle=\frac{\langle G|\hat{v}^s_{\k}(\tau)\hat{v}^{s'}_{\k'}(\tau)|G\rangle}{\langle G|G\rangle}.  
\end{equation}
It is important to note that for a general state $|G\rangle$, one-point function $\langle\hat{v}^s_{\k}(\tau)\rangle$ may not be zero even at late time ($\tau\rightarrow 0$) and as a result both $\langle\hat{\R}_{\k}(\tau)\rangle$ and $\langle\hat{h}^s_{\k}(\tau)\rangle$ can be nonzero. In that case, the power spectrum should be defined in the following way,
\begin{align}\label{physpowerspectrum}
\langle\hat{\O}^s_{\k}(\tau)\hat{\O}^{s'}_{\k'}(\tau)\rangle_{phy}\equiv&\langle(\hat{\O}^s_{\k}(\tau)-\langle\hat{\O}^s_{\k}(\tau)\rangle)(\hat{\O}^{s'}_{\k'}(\tau)-\langle\hat{\O}^{s'}_{\k'}(\tau)\rangle)\rangle\nonumber \\
=&\langle\hat{\O}^s_{\k}(\tau)\hat{\O}^{s'}_{\k'}(\tau)\rangle-\langle\hat{\O}^s_{\k}(\tau)\rangle\langle\hat{\O}^{s'}_{\k'}(\tau)\rangle.
\end{align}
With the initial state (\ref{Gstate}), we can calculate
\begin{align}
\langle\hat{v}^s_{\k}(\tau)\hat{v}^{s'}_{\k'}(\tau)\rangle=\frac{1}{2}(2 \pi)^3\delta^3(\k+\k')&|u_k(\tau)|^2 \delta_{ss'}+ A(s,s';\k, \k')u^*_k u^*_{k'}+ A^*(s,s';-\k, -\k')u_k u_{k'} \nonumber \\ 
&+ B(s,s';-\k, \k')u_k u^*_{k'}+B(s',s;-\k', \k)u^*_k u_{k'}, 
\end{align}
where,
\begin{equation}
A(s,s';\k, \k')=\frac{1}{2}\frac{\langle G|\hat{a}^s_{\k}\hat{a}^{s'}_{\k'}|G\rangle}{\langle G|G\rangle}, \qquad B(s,s';\k, \k')=\frac{1}{2}\frac{\langle G|\hat{a}^{s\dagger}_{\k}\hat{a}^{s'}_{\k'}|G\rangle}{\langle G|G\rangle}.
\end{equation}
And
\begin{eqnarray}
\langle\hat{v}^s_{\k}(\tau)\rangle\langle\hat{v}^{s'}_{\k'}(\tau)\rangle= b(s;\k)b(s';\k')u^*_k u^*_{k'} +b^*(s;-\k)b^*(s';-\k')u_k u_{k'}~~~~~~~~\nonumber \\ 
+ b^*(s;-\k)b(s';\k')u_k u^*_{k'}+b(s;\k)b^*(s';-\k')u^*_k u_{k'}, 
\end{eqnarray}
where,
\begin{equation}
b(s;\k)=\frac{1}{\sqrt{2}}\frac{\langle G|\hat{a}^s_{\k}|G\rangle}{\langle G|G\rangle}=\frac{1}{\sqrt{2}}\frac{\langle G_s|\hat{a}^s_{\k}|G_s\rangle}{\langle G_s|G_s\rangle}.
\end{equation}
Now, if $s\neq s'$, it can be shown very easily that $A(s,s'\neq s;\k, \k')=b(s;\k)b(s';\k')$ and $ B(s,s'\neq s;\k, \k')=b^*(s;\k)b(s';\k')$. And therefore,
\begin{equation}
\langle\hat{v}^s_{\k}(\tau)\hat{v}^{s'\neq s}_{\k'}(\tau)\rangle_{phy}=\langle\hat{v}^s_{\k}(\tau)\hat{v}^{s'\neq s}_{\k'}(\tau)\rangle-\langle\hat{v}^s_{\k}(\tau)\rangle\langle\hat{v}^{s'\neq s}_{\k'}(\tau)\rangle=0
\end{equation}
That leads to
\begin{equation}\label{physps}
\langle\hat{v}^s_{\k}(\tau)\hat{v}^{s'}_{\k'}(\tau)\rangle_{phy}=\langle\hat{v}^s_{\k}(\tau)\hat{v}^{s}_{\k'}(\tau)\rangle_{phy}\delta_{ss'}.
\end{equation}
Let us now calculate $\langle\hat{v}^s_{\k}(\tau)\hat{v}^{s'}_{\k'}(\tau)\rangle_{phy}$. Introducing $k_*=\sqrt{k k'}$, $\bar{k}=k+k'$ and $\Delta k=k-k'$ and using equation (\ref{modesBD}), we can write
\begin{align}
\langle\hat{v}^s_{\k}(\tau)\hat{v}^s_{\k'}(\tau)\rangle_{phy}=\frac{1}{2}(2\pi)^3\delta^3(\k+\k')\frac{1}{k}\left(1+\frac{1}{k^2 \tau^2}\right)+A_s(\k, \k')e^{-i \bar{k} \tau} \frac{1}{k_*}\left(1-\frac{i \bar{k}}{\tau k_*^2}-\frac{1}{k_*^2 \tau^2}\right)\nonumber \\
+A_s^*(-\k, -\k')e^{i \bar{k} \tau} \frac{1}{k_*}\left(1+\frac{i \bar{k}}{\tau k_*^2}-\frac{1}{k_*^2 \tau^2}\right)+B_s(-\k, \k')e^{i \Delta k \tau} \frac{1}{k_*}\left(1-\frac{i \Delta k}{\tau k_*^2}+\frac{1}{k_*^2 \tau^2}\right)\nonumber \\
+B_s(-\k', \k)e^{-i \Delta k \tau} \frac{1}{k_*}\left(1+\frac{i \Delta k}{\tau k_*^2}+\frac{1}{k_*^2 \tau^2}\right)\ ,
\end{align}
where,
\begin{equation}
A_s(\k, \k')=A(s,s;\k, \k')-b(s;\k)b(s;\k'), \quad B_s(\k, \k')=B(s,s;\k, \k')-b^*(s;\k)b(s;\k').
\end{equation}
In the superhorizon limit ($|k \tau|,|k' \tau|\ll 1$), we finally obtain
\begin{align}\label{finalpowerspec}
\langle\hat{v}^s_{\k}(\tau)\hat{v}^s_{\k'}(\tau)\rangle_{phy}\approx&\frac{1}{2}(2 \pi)^3\delta^3(\k+\k')\frac{1}{k}\left(1+\frac{1}{k^2 \tau^2}\right)+\left(\frac{1}{k_*^3 \tau^2}+ \frac{k^2+k'^2}{2 k_*^3}\right)\nonumber \\ 
\times&\left[-A_s(\k, \k')-A_s^*(-\k, -\k')+B_s(-\k, \k')+B_s(-\k', \k)\right] 
+\cdots 
\end{align}
where the dots indicate terms of higher order.

\subsection{Scalar power spectrum}
In the superhorizon limit, from the last equation at the leading order we obtain
\begin{align}\label{superhorizons}
\langle\hat{\R}_{\k}(\tau)\hat{\R}_{\k'}(\tau)\rangle_{phy}\approx\frac{1}{2} (2 \pi)^3 \frac{ H^4}{ \dot{\bar{\phi}}^2 k^3}\delta^3(\k+\k')
+\frac{ H^4}{ \dot{\bar{\phi}}^2 k_*^3}[&-A_0(\k, \k')-A_0^*(-\k, -\k') \nonumber\\
 &+B_0(-\k, \k')+B_0(-\k', \k)]. 
\end{align} 
Let us now simplify the last equation by making some assumptions about the initial state. Our universe as we see it today, is homogeneous and isotropic on large scale. Demanding homogeneity in the superhorizon limit restricts the form of the power spectrum 
\begin{equation}\label{twopoint}
\langle\hat{\R}_{\k}(\tau)\hat{\R}_{\k'}(\tau)\rangle_{phy}= P(\k, \tau)\delta^3(\k+\k')\delta_{ss'}. 
\end{equation}
Where $P(\k, \tau)$ is some arbitrary function of $\k$ and $ \tau$. If we also assume that the initial state is isotropic, then $P(\k, \tau)=P(k,\tau)$. Comparing the last equation with the leading order term of equation (\ref{superhorizons}), we also find that $P(k,\tau)$ does not depend on $\tau $ and hence $\langle\hat{\R}_{\k}(\tau)\hat{\R}_{\k'}(\tau)\rangle_{phy}$ is time-independent. All these assumptions about the initial state allow us to write
\begin{equation}\label{defw0}
 -A_0(\k, \k')-A_0^*(-\k, -\k')+B_0(-\k, \k')+B_0(-\k', \k)=(2 \pi)^3 W_0(k) \delta^3(\k+\k'),  
\end{equation}
where, $W_0(k)$ is some arbitrary function of $k$. Therefore the scalar power spectrum is given by
\begin{equation} 
\langle\hat{\R}_{\k}(\tau)\hat{\R}_{\k'}(\tau)\rangle= (2 \pi)^3 \frac{ H^4}{ \dot{\bar{\phi}}^2 k^3}\left(\frac{1}{2}+W_0(k)\right)\delta^3(\k+\k'),
\end{equation}and
\begin{align}
 \Delta^2_{\R}=\frac{ H^4}{4\pi^2~ \dot{\bar{\phi}}^2 }\left(1+2W_0(k)\right)\ , \qquad n_s=1-6\epsilon +2\eta +\frac{d \ln \left(1+2W_0(k)\right)}{d\ln k}
\end{align}
where, $W_0(k)$ is defined by equation (\ref{defw0}). Let us note that here we have assumed that energies of these states are not large enough to affect the slow-roll parameters.

\subsection{Tensor power spectrum}
For the tensor modes, in the superhorizon limit, at the leading order we have
\begin{align}\label{superhorizont}
\langle\hat{h}^s_{\k}(\tau)\hat{h}^{s'}_{\k'}(\tau)\rangle_{phy}\approx  (2 \pi)^3 \frac{ H^2}{ \M^2 k^3}\delta^3(\k+\k')\delta_{ss'}
+\frac{2 H^2}{\M^2 k_*^3}[&-A_s(\k, \k')-A_s^*(-\k, -\k') \nonumber\\
 &+B_s(-\k, \k')+B_s(-\k', \k)]\delta_{ss'}. 
\end{align} 
We can make similar assumptions about the initial state of tensor modes to obtain
\begin{equation}\label{secondcon}
 -A_s(\k, \k')-A_s^*(-\k, -\k')+B_s(-\k, \k')+B_s(-\k', \k)=(2 \pi)^3 W_s(k) \delta^3(\k+\k')  \ .
\end{equation}
Therefore, the tensor power spectrum is given by
\begin{equation} 
\langle\hat{h}^s_{\k}(\tau)\hat{h}^{s'}_{\k'}(\tau)\rangle_{phy}= (2 \pi)^3 \frac{ H^2}{ \M^2 k^3}\left(1+2W_s(k)\right)\delta^3(\k+\k')\delta_{ss'},
\end{equation}
and
\begin{align}
\Delta^2_{t}=\frac{ H^2}{\pi^2~ \M^2 }\left(1+\sum_{s=\times,+} W_s(k)\right)\ , \qquad n_t=-2\epsilon +\frac{d \ln \left(1+\sum_{s=\times,+} W_s(k)\right)}{d\ln k}
\end{align}
where, $W_s(k)$ is defined by equation (\ref{secondcon}).

We can further assume that $W_0(k)=W_+(k)=W_\times(k)$ which is reasonable because any pre-inflationary dynamics that excites the scalar modes will probably also excite the tensor modes in the same way. In that case, the tenser-to-scalar ratio remains unchanged 
\begin{equation}
r=16\epsilon \ .
\end{equation}
However, both $n_s$ and $n_t$ get corrected and hence the consistency relation $r=-8n_t$ is no longer true. Detection of primordial gravitational waves will provide us an important tool for probing the initial state.  

\

In the following sections, we will compute the three-point functions of scalar and tensor perturbations for slow-roll inflation with different initial states to check the validity of the non-Gaussianity consistency relations. But before we proceed, let us make some comments about renormalizability of the energy-momentum tensor of fluctuations. The initial state of  perturbations are not exactly free of constraints. It is reasonable to impose the constraint on the initial state $|G\rangle$ that it does not introduce any new ultra-violet divergences to the energy-momentum tensor. Therefore, to make sure that $\langle G|  \hat{\T}_{\mu \nu}|G \rangle$ has desired UV behavior, we can impose the following constraint on the initial state
\begin{equation}\label{emconstraint}
\langle G|  \hat{\T}_{\mu \nu}|G \rangle=\langle 0|  \hat{\T}_{\mu \nu}|0 \rangle+\text{UV finite}\ ,
\end{equation}
where, ${\T}_{\mu \nu}$ is the energy-momentum tensor of the perturbations.\footnote{It will be discussed in more details in section \ref{brn}. See \cite{Abramo:1997hu} for a more complete discussion.} The energy-momentum tensor of scalar perturbations ${\T}^s_{\mu \nu}$ can be found in \cite{Kundu:2011sg} and the energy-momentum tensor of tensor perturbations ${\T}^t_{\mu \nu}$ is given by
\begin{align}\label{emtensor}
8 \pi G \T^t_{00}=&\frac{\dot{a}}{a}\dot{h}_{kl}h_{kl}+\frac{1}{8}\left(\dot{h}_{kl}\dot{h}_{kl}+\frac{1}{a^2}\partial_m h_{kl}\partial_m h_{kl}\right),\\
8 \pi G \T^t_{ij}=& a^2 \delta_{ij}\left[\frac{3}{8 a^2}\partial_m h_{kl}\partial_m h_{kl}-\frac{3}{8}\dot{h}_{kl}\dot{h}_{kl}\right]+\frac{1}{2}a^2 \dot{h}_{ik}\dot{h}_{kj}+\frac{1}{4}\partial_i h_{kl}\partial_j h_{kl}-\frac{1}{2}\partial_l h_{ki}\partial_l h_{jk}.\nonumber
\end{align}

\section{Non-Gaussianities and general initial states}\label{ng}
In this section, we will set up the calculation for $\langle\hat{O}_{\k_1}(\tau)\hat{O}_{\k_2}(\tau)\hat{O}_{\k_3}(\tau)\rangle$ (where $\hat{O}_{\k}$ is either scalar perturbation $\hat{\R}_{\k}$ or tensor perturbation $\hat{h}^s_{\k}$) with a general initial state, then we will review the results for the Bunch-Davies state. For a general state $|G\rangle$, the operator $\hat{O}_{\k}$ can have a non vanishing expectation value and hence physically relevant part of the three-point function is given by,
\begin{equation}\label{3ptfn}
\langle\hat{O}_{\k_1}(\tau)\hat{O}_{\k_2}(\tau)\hat{O}_{\k_3}(\tau)\rangle_{phy}\equiv \langle \{\hat{O}_{\k_1}(\tau)-\langle \hat{O}_{\k_1}(\tau) \rangle \}\{\hat{O}_{\k_2}(\tau)-\langle \hat{O}_{\k_2} (\tau)\rangle \}\{\hat{O}_{\k_3}(\tau)-\langle \hat{O}_{\k_3} (\tau)\rangle \}\rangle\ ,
\end{equation}
where $\langle \hat{A}\rangle\equiv\langle G|\hat{A}|G\rangle$. Now, using time-dependent perturbation theory, for any operator (e.g. $\hat{O}_{\k_1}(\tau)\hat{O}_{\k_2}(\tau)...$) we have,
\begin{align}
\langle G|\hat{O}_{\k_1}(\tau)\hat{O}_{\k_2}(\tau)...|G\rangle=\langle G|\left(\bar{T} e^{i \int_{\tau_0}^{\tau}H^{I}_{int}(\tau')d\tau'} \right) \hat{O}_{\k_1}^{I}(\tau)\hat{O}_{\k_2}^{I}(\tau)...\left(T e^{-i \int_{\tau_0}^{\tau}H^{I}_{int}(\tau')d\tau'} \right)|G\rangle, 
\end{align}
where all fields are in the interaction picture and $H^I_{int}(\tau)$ is the interacting part of the Hamiltonian in the interaction picture. $T$ and $\bar{T}$ are the time and anti-time ordered product respectively. $\tau_0$ is the conformal time at the beginning of inflation. At first order in perturbation theory, we obtain,
\begin{align}\label{npoint}
\langle G &|\hat{O}_{\k_1}(\tau)\hat{O}_{\k_2}(\tau)...|G\rangle\nonumber\\
&= \langle G|\hat{O}_{\k_1}^{I}(\tau)\hat{O}_{\k_2}^{I}(\tau)...|G\rangle-i\int_{\tau_0}^{\tau}d\tau'\langle G|\left[\hat{O}_{\k_1}^{I}(\tau)\hat{O}_{\k_2}^{I}(\tau)..., H^{I}_{int}(\tau')\right]|G\rangle.
\end{align}
So far our discussion is very general and does not depend on the details of the inflationary model. 

For slow-roll inflation, we can use equations (\ref{3ptfn},\ref{npoint}) to calculate different three-point functions. We are interested in the late time behavior of the three-point functions i.e we will take the usual limit $\tau\rightarrow 0$. For simplicity we will assume that for the free theory, the operator expectation value $\langle \hat{O}_{\k}(\tau)\rangle$ in state $|G\rangle$ vanishes at late time.\footnote{We should note that interactions can generate non vanishing one-point functions  even for these states.} Therefore, at first order in slow-roll parameters, three-point functions are given by,
\begin{align}\label{effthreepointfun}
\langle G|\hat{O}_{\k_1}(\tau)\hat{O}_{\k_2}(\tau)\hat{O}_{\k_3}(\tau)|G\rangle_{phy}=&\langle G|\hat{O}_{\k_1}(\tau)\hat{O}_{\k_2}(\tau)\hat{O}_{\k_3}(\tau)|G\rangle \nonumber \\
&-\left(\langle G|\hat{O}_{\k_1}(\tau)|G\rangle \langle G|\hat{O}_{\k_2}(\tau)\hat{O}_{\k_3}(\tau)|G\rangle+ \text{cyclic perm}\right),
\end{align}
where all the quantities in the right hand side should be computed using equation (\ref{npoint}).

\subsection{Bunch-Davies state}
Next, as a warm up exercise, we will calculate the three-point functions for the Bunch-Davies initial state 
\begin{equation}
|0\rangle \equiv |0\rangle_{\text{scalar}}\varotimes|0_{s=+}\rangle\varotimes |0_{s=\times}\rangle
\end{equation}
to demonstrate that all the consistency relations (\ref{cr},\ref{F}) are satisfied and in the leading order we obtain
\begin{equation}
\F\equiv
 \begin{pmatrix}
  \F_{\R\R}&~~~~~  \F_{\R h}  \\
  \F_{h\R} &~~~~ \F_{hh} 
 \end{pmatrix}
 =
  \begin{pmatrix}
  -6\epsilon+2\eta ~~~& 3/2  \\
  -2\epsilon ~~& 3/2
 \end{pmatrix}\ .
\end{equation}
\subsubsection{Three scalars correlator}\label{bd3pt}
At leading order in the slow-roll parameters, the third order action for scalar fluctuations is given by \cite{Maldacena:2002vr}
\begin{equation}\label{thirdorder}
S_3= -8 \pi G \int d^3x d\tau ~ a^3(\tau)\left( \frac{\dot{\bar{\phi}}}{H}\right)^4 H \R_c'^2 \partial ^{-2}\R_c',
\end{equation}
where, $\R_c$ is the redefined field
\begin{equation}\label{rc1}
\R=\R_c -\frac{1}{4} \left(3\epsilon -2\eta\right)\R_c^2- \frac{1}{2} \epsilon ~ \partial ^{-2}\left(\R_c \partial^2 \R_c\right).
\end{equation}
In momentum space the last equation becomes
\begin{equation}\label{rc}
\R_{\k}=\R_{c,\k}-\frac{1}{4}\left(3\epsilon -2\eta\right)\int \frac{d^3\mathbf{\p}}{(2\pi)^{3}}\R_{c,\p}\R_{c,\k-\p}-\frac{1}{2} \epsilon ~\int \frac{d^3\mathbf{\p}}{(2\pi)^{3}}\frac{(\k-\p)^2}{k^2}\R_{c,\p}\R_{c,\k-\p}.
\end{equation}
The interaction Hamiltonian can be found from $S_3=- \int d\tau H_{int}$. In momentum space $H_{int}$ is given by
\begin{equation}\label{Hint}
H_{int}(\tau)=-\frac{8 \pi G}{(2\pi)^6}  a^3(\tau)\left( \frac{\dot{\bar{\phi}}}{H}\right)^4 H \int d^3\p_1 d^3\p_2 d^3\p_3 \left(\frac{1}{p_3^2}\right) \R'_{\p_1}(\tau)\R'_{\p_2}(\tau)\R'_{\p_3}(\tau) \delta^3(\p_1+\p_2+\p_3).
\end{equation}
For the Bunch-Davies state, only the first term of equation (\ref{effthreepointfun}) contributes and we obtain,
\begin{align}\label{3sbd}
\langle 0|\hat{\R}_{\k_1}(\tau)\hat{\R}_{\k_2}(\tau)\hat{\R}_{\k_3}(\tau)|0\rangle_{phy}=&\langle 0|\hat{\R}_{\k_1}^{I}(\tau)\hat{\R}_{\k_2}^{I}(\tau)\hat{\R}_{\k_3}^{I}(\tau)|0\rangle\nonumber\\
&-i\int_{\tau_0}^{\tau}d\tau'\langle 0|\left[\hat{\R}_{\k_1}^{I}(\tau)\hat{\R}_{\k_2}^{I}(\tau)\hat{\R}_{\k_3}^{I}(\tau), H^{I}_{int}(\tau')\right]|0\rangle.
\end{align}
The first term in the last equation can be written using the redefined field (\ref{rc})
\begin{align}\label{pert}
\langle 0|\hat{\R}_{\k_1}^{I}(\tau)&\hat{\R}_{\k_2}^{I}(\tau)\hat{\R}_{\k_3}^{I}(\tau)|0\rangle= &\\
- &\frac{1}{4}\left(3\epsilon -2\eta\right)\left(\int \frac{d^3\mathbf{\p}}{(2\pi)^{3}}\langle 0|\hat{\R}_{c,\k_1}^{I}(\tau)\hat{\R}_{c,\k_2}^{I}(\tau)\hat{\R}_{c,\p}^{I}(\tau)\hat{\R}_{c,\k_3-\p}^{I}(\tau)|0\rangle+\text{cyclic perm}\right) \nonumber \\
- &\frac{1}{2} \epsilon\left(\int \frac{d^3\mathbf{\p}}{(2\pi)^{3}}\frac{(\k_3-\p)^2}{k_3^2}\langle 0|\hat{\R}_{c,\k_1}^{I}(\tau)\hat{\R}_{c,\k_2}^{I}(\tau)\hat{\R}_{c,\p}^{I}(\tau)\hat{\R}_{c,\k_3-\p}^{I}(\tau)|0\rangle+\text{cyclic perm}\right). \nonumber
\end{align}
$\hat{\R}_{c,\k}^{I}(\tau)$ behaves like the free field, and can be written as
\begin{equation}
\hat{\R}_{c,\k}^{I}(\tau)=\frac{1}{\sqrt{2}}\left[\hat{a}^0_{\k}\R^*_k(\tau)+ \hat{a}^{0\dagger}_{-\k}\R_k(\tau)\right],
\end{equation}
where $\R_k(\tau)=\left(\frac{H}{a \dot{\bar{\phi}}} \right) \frac{e^{ik\tau}}{\sqrt{k}}\left(1+\frac{i}{k\tau}\right)$ and operator $\hat{a}^0_{\k}$ annihilates $|0\rangle_{\text{scalar}}$. At the leading order in the slow-roll parameters equation (\ref{pert}) becomes
\begin{align}
\langle 0|\hat{\R}_{\k_1}^{I}(\tau)\hat{\R}_{\k_2}^{I}(\tau)\hat{\R}_{\k_3}^{I}(\tau)|0\rangle= &-(2 \pi)^3 \delta^3(\k_1+\k_2+\k_3)P_R(k_2)P_R(k_1)\\
&\times\left[\frac{1}{2}\left(3\epsilon -2\eta+\epsilon\frac{k_1^2+k_2^2}{k_3^2}\right)\right]+ \text{cyclic perm}.\nonumber
\end{align}
Next term in the equation (\ref{3sbd}) can be easily computed yielding\footnote{For large $\tau_0$, all exponentials with $\tau_0$ will oscillate. When performing the calculations, we can either use the average value (i.e. zero) for them or we can choose an integration contour such that the oscillating pieces decrease exponentially for large $\tau_0$ \cite{Maldacena:2002vr}.}
\begin{align}
-i\int_{\tau_0}^{\tau}d\tau'\langle 0|\left[\hat{\R}_{\k_1}^{I}(\tau)\hat{\R}_{\k_2}^{I}(\tau)\hat{\R}_{\k_3}^{I}(\tau), H^{I}_{int}(\tau')\right]|0\rangle&=-(2 \pi)^3 \delta^3(\k_1+\k_2+\k_3)P_R(k_2)P_R(k_1)\nonumber \\
&\times \left(\frac{4\epsilon }{ (k_1+k_2+k_3)}\frac{k_1^2k_2^2}{k_3^3}\right)+ \text{cyclic perm}.
\end{align}
Therefore, finally we have,
\begin{align}\label{sssnongauss}
\langle 0|\hat{\R}_{\k_1}(\tau)\hat{\R}_{\k_2}&(\tau)\hat{\R}_{\k_3}(\tau)|0\rangle_{phy}=-(2 \pi)^3 \delta^3(\sum\k)P_R(k_2)P_R(k_1)\nonumber \\
&\times \left[\frac{1}{2}\left(3\epsilon -2\eta+\epsilon\frac{k_1^2+k_2^2}{k_3^2}\right)+\frac{4\epsilon }{ (k_1+k_2+k_3)}\frac{k_1^2k_2^2}{k_3^3}\right]+ \text{cyclic perm} ,
\end{align}
where $\sum\k=\k_1+\k_2+\k_3$. Therefore, in the squeezed limit, $f^{loc}_{NL}$ is given by,
\begin{equation}
f_{NL}^{loc}\approx \frac{5}{12}(1-n_s).
\end{equation}
\subsubsection{Two scalars and a graviton correlator}
At leading order in the slow-roll parameters, the relevant part of the action is given by \cite{Maldacena:2002vr}
\begin{equation}\label{actionssh}
S_3= \frac{1}{2} \int d^3x d\tau ~ a(\tau)^2\left( \frac{\dot{\bar{\phi}}^2}{H^2}\right) h_{ij} \partial_i \R_c \partial_j\R_c\ ,
\end{equation}
where, $\R_c$ is again a redefined field which has a form similar to (\ref{rc1}), however, for this computation only the leading part is important and hence $\R_c=\R$.

In momentum space $H_{int}$ is given by
\begin{align}\label{Hssh}
H_{int}(\tau)=\frac{a^2(\tau)}{2(2\pi)^6}\left( \frac{\dot{\bar{\phi}}^2}{H^2}\right)\sum_{s'=+,\times}\int & d^3\p_1 d^3\p_2 d^3\p_3   \epsilon^{s'}_{ij}(\p_3)\p_{1,i}\p_{2,j}\nonumber\\
&\times\R_{\p_1}(\tau)\R_{\p_2}(\tau) h^{s'}_{\mathbf{p_3}}(\tau) \delta^3(\p_1+\p_2+\p_3).
\end{align}
And hence at first order in perturbation theory, the three-point function is given by
\begin{align}
\langle 0|\hat{\R}_{\k_1}(\tau)\hat{\R}_{\k_2}(\tau)\hat{h}^s_{\k_3}(\tau)|0\rangle_{phy}=&\langle 0|\hat{\R}_{\k_1}^{I}(\tau)\hat{\R}_{\k_2}^{I}(\tau)\hat{h}^{s,I}_{\k_3}(\tau)|0\rangle\nonumber\\
&-i\int_{\tau_0}^{\tau}d\tau'\langle 0|\left[\hat{\R}_{\k_1}^{I}(\tau)\hat{\R}_{\k_2}^{I}(\tau)\hat{h}^{s,I}_{\k_3}(\tau), H^{I}_{int}(\tau')\right]|0\rangle.
\end{align}
The first term in the last equation vanishes. The second term can be computed easily, yielding 
\begin{equation}
 \langle 0|\hat{\R}_{\k_1}(\tau)\hat{\R}_{\k_2}(\tau)\hat{h}^s_{\k_3}(\tau)|0\rangle_{phy}=(2\pi)^3 \delta^3(\sum\k)\frac{H^6}{2\M^2 \dot{\bar{\phi}}^2} \epsilon^{s}_{ij}(\k_3)\k_{1,i}\k_{2,j}\I(k_1,k_2,k_3)
\end{equation}
where,
\begin{equation}
\I(k_1,k_2,k_3)= \frac{1}{(k_1k_2k_3)^3}\left(-k_t+\frac{k_1k_2k_3}{k_t^2}+\frac{k_1k_2+k_2k_3+k_1k_3}{k_t}\right)
\end{equation}
with $k_t=k_1+k_2+k_3$.

Let us consider two limiting cases. In the limit $k_3<<k_1,k_2$, we recover
\begin{equation}
 \langle 0|\hat{\R}_{\k_1}(\tau)\hat{\R}_{\k_2}(\tau)\hat{h}^s_{\k_3}(\tau)|0\rangle_{phy}=(2\pi)^3 \delta^3(\sum\k)P_\R(k_1) P_h(k_3) \frac{\epsilon^{s}_{ij}(\k_3)\k_{1,i}\k_{1,j}}{k_1^2}\left(\frac{3}{2}\right)\ .
\end{equation}
Where we have used the fact that $\epsilon^{s}_{ij}(\k_3)\k_{1,i}\k_{2,j}=-\epsilon^{s}_{ij}(\k_3)\k_{1,i}\k_{1,j}$. Recall that $n_s\sim 1$ and hence this result is consistent with (\ref{cr}) and (\ref{F}). 

Also note that in the limit $k_3<<k_1,k_2$, we get
\begin{equation}
 \frac{\langle 0|\hat{\R}_{\k_1}(\tau)\hat{h}^s_{\k_2}(\tau)\hat{\R}_{\k_3}(\tau)|0\rangle_{phy}}{P_h(k_1) P_\R(k_3)}\approx\O\left(\frac{k_3^2}{k_1^2}\right)
\end{equation}
and hence consistent with (\ref{cr}).

\subsubsection{Two gravitons and a scalar correlator}
At leading order in the slow-roll parameters, the relevant part of the action is given by 
\begin{equation}\label{actionssh}
S_3=- \frac{1}{4} \int d^3x d\tau ~ a(\tau)^3H\left( \frac{\dot{\bar{\phi}}^2}{H^2}\right) h'_{ij}h'_{ij} \partial^{-2}\R'_c \ ,
\end{equation}
where, following \cite{Maldacena:2002vr} we have done further field redefinition 
\begin{equation}
\R=\R_c+\frac{1}{32}h_{ij}h_{ij}-\frac{1}{16}\partial^{-2}\left(h_{ij}\partial^2 h_{ij}\right)+...
\end{equation}
where dots represent terms that are negligible outside the horizon. In momentum space the last equation becomes,
\begin{equation}
\R_{\k}=\R_{c,\k}+\frac{1}{32} \sum_{s,s'=+,\times}\int \frac{d^3\mathbf{\p}}{(2\pi)^{3}}h^s_{\p}h^{s'}_{\k-\p}\epsilon^s_{ij}(\p)\epsilon^{s'}_{ij}(\k-\p)\left(1-\frac{2(\k-\p)^2}{k^2}\right)\ .
\end{equation}
In momentum space, the interaction Hamiltonian is given by
\begin{align}\label{Hhhs}
H_{int}(\tau)=- \frac{1}{4(2\pi)^6} a^3(\tau)H\left( \frac{\dot{\bar{\phi}}^2}{H^2}\right) &\sum_{s_1,s_2=+,\times}\int d^3\p_1 d^3\p_2 d^3\p_3 \left(\frac{1}{p_3^2}\right) \\
&\times {h^{s_1}_{\p_1}}'(\tau){h^{s_2}_{\p_2}}'(\tau)\R'_{c,\p_3}(\tau)\epsilon^{s_1}_{ij}(\p_1)\epsilon^{s_2}_{ij}(\p_2)\delta^3(\p_1+\p_2+\p_3)\ . \nonumber
\end{align}
Hence at first order in perturbation theory, the three-point function is given by
\begin{align}
\langle 0|\hat{h}^{s}_{\k_1}(\tau)\hat{h}^{s'}_{\k_2}(\tau)\hat{\R}_{\k_3}(\tau)|0\rangle_{phy}=&\langle 0|\hat{h}^{s,I}_{\k_1}(\tau)\hat{h}^{s',I}_{\k_2}(\tau)\hat{\R}^I_{\k_3}(\tau)|0\rangle\\
&-i\int_{\tau_0}^{\tau}d\tau'\langle 0|\left[\hat{h}^{s,I}_{\k_1}(\tau)\hat{h}^{s',I}_{\k_2}(\tau)\hat{\R}^I_{\k_3}(\tau), H^{I}_{int}(\tau')\right]|0\rangle.\nonumber
\end{align}
The first term of the last equation is nonzero
\begin{equation}
\langle 0|\hat{h}^{s,I}_{\k_1}(\tau)\hat{h}^{s',I}_{\k_2}(\tau)\hat{\R}^I_{\k_3}(\tau)|0\rangle=\frac{(2\pi)^3}{16}P_h(k_1)P_{h}(k_2)\epsilon^{s}_{ij}(\k_1)\epsilon^{s'}_{ij}(\k_2)\left(\frac{k_3^2-k_1^2-k_2^2}{k_3^2}\right)\delta^3(\sum\k)\ .
\end{equation}
The second term can also be computed using the interaction Hamiltonian (\ref{Hhhs}), yielding
\begin{align}
-i\int_{\tau_0}^{\tau}d\tau'\langle 0|\left[\hat{h}^{s,I}_{\k_1}(\tau)\hat{h}^{s',I}_{\k_2}(\tau)\hat{\R}^I_{\k_3}(\tau), H^{I}_{int}(\tau')\right]|0\rangle=-&\frac{(2\pi)^3}{2} P_h(k_1)P_{h}(k_2)\epsilon^{s}_{ij}(\k_1)\epsilon^{s'}_{ij}(\k_2) \nonumber\\
&\times \left(\frac{k_1^2k_2^2}{k_3^3 k_t}\right)\delta^3(\sum\k)\ .
\end{align}
Therefore,
\begin{align}
\langle 0|\hat{h}^{s}_{\k_1}(\tau)\hat{h}^{s'}_{\k_2}(\tau)\hat{\R}_{\k_3}(\tau)|0\rangle_{phy}&=\frac{(2\pi)^3}{2} P_h(k_1)P_{h}(k_2)\epsilon^{s}_{ij}(\k_1)\epsilon^{s'}_{ij}(\k_2)\\
&~~\times \left[\frac{1}{8}\left(\frac{k_3^2-k_1^2-k_2^2}{k_3^2}\right)-\left(\frac{k_1^2k_2^2}{k_3^3 k_t}\right)\right]\delta^3(\sum\k)\ . \nonumber
\end{align}
In the limit $k_3<<k_1=k_2$, we obtain
\begin{align}
\langle 0|\hat{h}^{s}_{\k_1}(\tau)\hat{h}^{s'}_{\k_2}(\tau)\hat{\R}_{\k_3}(\tau)|0\rangle_{phy}&\approx-\frac{(2\pi)^3}{2} P_h(k_1)P_{h}(k_1)\delta_{ss'}\left(\frac{k_1^3}{k_3^3}\right)\delta^3(\sum\k)\nonumber\\
&=(2\pi)^3\ P_h(k_1)P_{\R}(k_3)n_t\delta_{ss'}\delta^3(\sum\k)\ ,
\end{align}
where we have used the fact that $n_t=-2\epsilon$. 

Note that in the other squeezed limit $k_3<<k_1=k_2$, we obtain
\begin{align}
\frac{\langle 0|\hat{h}^{s}_{\k_1}(\tau)\hat{\R}_{\k_2}(\tau)\hat{h}^{s'}_{\k_3}(\tau)|0\rangle_{phy}}{P_h(k_3)P_{\R}(k_2)}\approx  \O\left( \frac{k_3^2}{k_1^2} \epsilon\right) \ .
\end{align}
Therefore, two gravitons and a scalar three-point functions in the squeezed limit agree with the consistency conditions (\ref{cr}) and (\ref{F}).

\subsubsection{Three gravitons correlator}
The third order action for the 3-gravitons interaction is given by\footnote{For detailed discussions see \cite{Maldacena:2002vr,Shiraishi:2011st,Soda:2011am,Maldacena:2011nz}.}
\begin{equation}\label{actionhhh}
S_3= \frac{\M^2}{4} \int d^3x d\tau ~ a(\tau)^2\left( h_{ik}h_{jl}-\frac{1}{2}h_{ij}h_{kl}\right)\partial_k \partial_l h_{ij} \ ,
\end{equation}
The interaction Hamiltonian can be found from $S_3=- \int d\tau H_{int}$. In momentum space $H_{int}$ is given by
\begin{equation}\label{Hhint}
H_{int}(\tau)=\frac{\M^2a^2(\tau)}{4(2\pi)^6}   \int d^3\p_1 d^3\p_2 d^3\p_3 \sum_{s_1,s_2,s_3}h^{s_1}_{\p_1}(\tau) h^{s_2}_{\p_2}(\tau) h^{s_3}_{\p_3}(\tau)T(\p_1,\p_2,\p_3; s_1,s_2,s_3) \delta^3\left(\sum\p\right)\ ,
\end{equation}
where, 
\begin{equation}
T(\p_1,\p_2,\p_3; s_1,s_2,s_3)=\left(\epsilon^{s_1}_{ik}(\p_1)\epsilon^{s_2}_{jl}(\p_2)-\frac{1}{2}\epsilon^{s_1}_{ij}(\p_1)\epsilon^{s_2}_{kl}(\p_2)\right)\epsilon^{s_3}_{ij}(\p_3)\p_{3,k}\p_{3,l}\ .
\end{equation}
At first order in the perturbation theory, we obtain
\begin{align}
\langle 0|\hat{h}^s_{\k_1}(\tau) \hat{h}^{s'}_{\k_2}(\tau)\hat{h}^{s''}_{\k_3}(\tau)|0\rangle_{phy}=&\langle 0|\hat{h}^{s,I}_{\k_1}(\tau) \hat{h}^{s',I}_{\k_2}(\tau)\hat{h}^{s'',I}_{\k_3}(\tau)|0\rangle\nonumber\\
&-i\int_{\tau_0}^{\tau}d\tau'\langle 0|\left[\hat{h}^{s,I}_{\k_1}(\tau) \hat{h}^{s',I}_{\k_2}(\tau)\hat{h}^{s'',I}_{\k_3}(\tau), H^{I}_{int}(\tau')\right]|0\rangle.
\end{align}
The first term in the last equation vanishes. The second term is nonzero and the final result is
\begin{align}\label{3hbd}
\langle 0|\hat{h}^s_{\k_1}(\tau) \hat{h}^{s'}_{\k_2}(\tau)\hat{h}^{s''}_{\k_3}(\tau)|0\rangle_{phy}=&(2\pi)^3 \delta^3(\k_1+\k_2+\k_3)\frac{H^4}{2\M^4 }\I(k_1,k_2,k_3)\\
&\times \left[T(\k_1,\k_2,\k_3; s,s',s'')+ \text{all permutations}\right]\ , \nonumber
\end{align}
where,
\begin{equation}
\I(k_1,k_2,k_3)= \frac{1}{(k_1k_2k_3)^3}\left(-k_t+\frac{k_1k_2k_3}{k_t^2}+\frac{k_1k_2+k_2k_3+k_1k_3}{k_t}\right)
\end{equation}
with $k_t=k_1+k_2+k_3$. The three-point function (\ref{3hbd}) can be simplified further (see \cite{Maldacena:2002vr})
\begin{align}
\langle 0|\hat{h}^s_{\k_1}(\tau) \hat{h}^{s'}_{\k_2}(\tau)\hat{h}^{s''}_{\k_3}(\tau)|0\rangle_{phy}=(2\pi)^3 \delta^3(\sum\k)\frac{H^4}{2\M^4 }\I(k_1,k_2,k_3)~~~~~~~~~~\\
\times \left(-\epsilon^{s}_{ii'}(\k_1)\epsilon^{s'}_{jj'}(\k_2)\epsilon^{s''}_{ll'}(\k_3)t_{ijl}t_{i'j'l'}\right)\ , \nonumber
\end{align}
where,
\begin{equation}
t_{ijl}=\k_{1,l}\delta_{ij}+\k_{2,i}\delta_{jl}+\k_{3,j}\delta_{il}\ .
\end{equation}
In the squeezed limit $k_3<<k_1,k_2$, we obtain,
\begin{align}
\langle 0|\hat{h}^s_{\k_1}(\tau) \hat{h}^{s'}_{\k_2}(\tau)\hat{h}^{s''}_{\k_3}(\tau)|0\rangle_{phy}=(2\pi)^3 \delta^3(\sum \k)P_h(k_1)P_h(k_3)\delta_{ss'}\left(\frac{3}{2}\right)\frac{\k_{1;i} \k_{1;j}\epsilon^{s''}_{ij}(\k_3)}{ k_1^2}\ .
\end{align}

Therefore, all the three-point functions of scalar and tensor perturbations for slow-roll inflation with the Bunch-Davies initial state are consistent with the consistency relations (\ref{cr},\ref{F}) and the non-Gaussianity matrix $\F$ is given by
\begin{equation}
\F\equiv
 \begin{pmatrix}
  \F_{\R\R}&~~~~~  \F_{\R h}  \\
  \F_{h\R} &~~~~ \F_{hh} 
 \end{pmatrix}
 =
  \begin{pmatrix}
  -6\epsilon+2\eta ~~~& 3/2  \\
  -2\epsilon ~~& 3/2
 \end{pmatrix}\ .
\end{equation}
\subsection{Non-Gaussianities from coherent states}
Now we will calculate the three-point functions for slow-roll inflation with a non-Bunch-Davies initial state of fluctuations. First we will consider coherent states, defined as
\begin{equation}
\hat{a}^{s}_{\k}|C\rangle = C(\k;s)|C\rangle\ , \qquad s=0\ ,+\ , \times\ ,
\end{equation}
where again $s=0$ corresponds to scalar perturbations. Without loss of generality, we can impose the restriction that $\langle\hat{\R}_\k(\tau)\rangle=\langle\hat{h}^s_\k(\tau)\rangle=0$, in the superhorizon limit (before we introduce three point interactions). That leads to the condition 
\begin{equation}\label{cohcon}
C^*(-\k;s)=C(\k;s).
\end{equation}
However, we will show that interactions will generator non vanishing one-point functions even for these states. The functions $C(\k;s)$ are not entirely free of constraints; these states will not introduce any new UV-divergences to the energy-momentum tensor only if $C(\k;s)$ goes to zero faster than $\frac{1}{k^{5/2}}$ for large $\k$.

Coherent state is a special state because it closely resembles classical harmonic oscillation. We do not know anything about the physics before inflation, a priori any excited state is as good an initial state as the Bunch-Davies state. In particular, it has been shown explicitly in \cite{Kundu:2011sg} that at late time, three scalar bispectrum for coherent initial state is identical to that with the Bunch-Davies initial state (\ref{sssnongauss}). In this section we will argue that the same is true for all the three-point functions, i.e., in the limit $\tau\rightarrow 0$,
\begin{equation}\label{ngcoh}
\langle C|\hat{O}_{1;\k_1}(\tau)\hat{O}_{2;\k_2}(\tau)\hat{O}_{3;\k_3}(\tau)|C\rangle_{phy}=\langle 0|\hat{O}_{1;\k_1}(\tau)\hat{O}_{2;\k_2}(\tau)\hat{O}_{3;\k_3}(\tau)|0\rangle_{phy}
\end{equation}
where $\hat{O}_{n;\k_n}(\tau)$, (with $ n=1,2,3...$) are either scalar perturbation $\hat{\R}_{\k}$ or tensor perturbation $\hat{h}^s_{\k}$. It is not very difficult to understand why that is the case. One can think of coherent state as zero-point quantum fluctuations around some classical state. So, the field $\hat{O}^{coh}_{\k}(\tau)$ in the coherent state can be written as $\hat{O}^{coh}_{\k}(\tau)=O^{cl}(\tau) +\hat{O}^{vac}_{\k}(\tau)$, where classical part $O^{cl}(\tau)$ is obviously the expectation value $\langle\hat{O}^{coh}_{\k}(\tau)\rangle$; and $\hat{O}^{vac}_{\k}(\tau)$ is the original quantum field but now in the vacuum state. Interactions will generate non-zero one-point function even at late time, however, that will contribute only to the classical part. Only the quantum fluctuations contribute to the physically relevant part of three-point correlations and hence they remain unchanged (\ref{ngcoh}). Let us now make this discussion precise by performing a tree-level computation.

Before we proceed,  let us note few things. First of all, a coherent state is annihilated by an operator $\hat{c}^s_\k$,
\begin{equation}
\hat{c}^{s}_{\k}|C\rangle = 0\ , \qquad \text{where,} \qquad \hat{c}^{s}_{\k}=\hat{a}^{s}_{\k}- C(\k;s)
\end{equation}
and 
\begin{equation}
\left[{\hat{c}^s}_{\k_1},\hat{c}^{s'\dagger}_{\k_2}\right]=(2 \pi)^3 \delta^3(\k_1-\k_2)\delta_{s s'}\ , \qquad s,s'=0\ ,+\ , \times\ .
\end{equation}
Any operator $\hat{O}_{n;\k_n}(\tau)$ of the free theory can be written in terms of operators ${\hat{c}^s}_{\k_n}$ and $\hat{c}^{s\dagger}_{-\k_n}$
\begin{equation}\label{free}
\hat{O}_{n;\k_n}(\tau)=\frac{1}{\sqrt{2}}\left[{\hat{c}^s}_{\k_n}u^*_{n; k_n}(\tau)+\hat{c}^{s\dagger}_{-\k_n}u_{n;k_n}(\tau)\right]+\bar{O}_{n;\k_n}(\tau)\ ,
\end{equation}
where $u_{n;k_n}(\tau)$ is the mode function associated with the operator $\hat{O}_{n;\k_n}(\tau)$ and for slow-roll inflation $u_{n;k_n}(\tau)= \frac{e^{ik\tau}}{\sqrt{k}}\left(1+\frac{i}{k\tau}\right)$ (up to a factor which is not important for our purpose). $\bar{O}_{n;\k_n}(\tau)$ is the classical part 
\begin{equation}
\bar{O}_{n;\k_n}(\tau)=\langle C|\hat{O}_{n;\k_n}(\tau)|C\rangle= \frac{1}{\sqrt{2}}\left[C(\k_n)u^*_{n; k_n}(\tau)+C^*(-\k_n)u_{n;k_n}(\tau)\right]\ .
\end{equation}
Now one can easily show that for the free theory
\begin{equation}\label{eqn1}
\langle C|\hat{O}_{1;\k_1}(\tau_1)\hat{O}_{2;\k_2}(\tau_2)...|C\rangle=\langle 0|\left(\hat{O}_{1;\k_1}(\tau_1)+\bar{O}_{1;\k_1}(\tau_1)\right)\left(\hat{O}_{2;\k_2}(\tau_2)+\bar{O}_{2;\k_2}(\tau_2)\right)...|0\rangle\ .
\end{equation}
Note that at late time ($\tau\rightarrow 0$), $\bar{O}_{n;\k_n}(\tau)=0$ because of the condition (\ref{cohcon}). Now let us turn on interactions and compute the three-point function $\langle C|\hat{O}_{1;\k_1}(\tau)\hat{O}_{2;\k_2}(\tau)\hat{O}_{3;\k_3}(\tau)|C\rangle$ in the limit $\tau\rightarrow 0$. In first order in perturbation theory, we obtain
\begin{align}\label{eqn2}
\langle C|\hat{O}_{1;\k_1}(\tau)\hat{O}_{2;\k_2}(\tau)\hat{O}_{3;\k_3}(\tau)|C\rangle=&\langle C|\hat{O}^I_{1;\k_1}(\tau)\hat{O}^I_{2;\k_2}(\tau)\hat{O}^I_{3;\k_3}(\tau)|C\rangle\nonumber\\
&-i\int_{\tau_0}^{\tau}d\tau'\langle C|\left[\hat{O}^I_{1;\k_1}(\tau)\hat{O}^I_{2;\k_2}(\tau)\hat{O}^I_{3;\k_3}(\tau), H^{I}_{int}(\tau')\right]|C\rangle.
\end{align}
All the fields are now in the interaction picture. In the interaction picture fields behave like free fields and hence can be written in the form (\ref{free}). The first term in the last equation is evaluated at time $\tau\rightarrow 0$ and hence from equation (\ref{eqn1}) we get
\begin{equation}
\langle C|\hat{O}^I_{1;\k_1}(\tau)\hat{O}^I_{2;\k_2}(\tau)\hat{O}^I_{3;\k_3}(\tau)|C\rangle=\langle 0|\hat{O}^I_{1;\k_1}(\tau)\hat{O}^I_{2;\k_2}(\tau)\hat{O}^I_{3;\k_3}(\tau)|0\rangle\ .
\end{equation}
Before we proceed further, a few comments are in order: one can naively assume that the quantity $\langle 0|\hat{O}^I_{1;\k_1}(\tau)\hat{O}^I_{2;\k_2}(\tau)\hat{O}^I_{3;\k_3}(\tau)|0\rangle$ vanishes. However, it is important to note that this quantity can be non-zero because some of the relevant three-point interactions are written in terms of redefined fields which generally have a quadratic piece (see section \ref{bd3pt} for an example). 

The second term in equation (\ref{eqn2}) is more complicated because it depends on the full history.  The interaction Hamiltonian in momentum space, for the cases we are interested in, can be written in the following form
\begin{equation}
H_{int}(\tau')=\lambda(\tau') \int d^3\p_1 d^3\p_2 d^3\p_3 f(\p_1,\p_2,\p_3) \hat{M}_{1';\p_1}(\tau')\hat{M}_{2';\p_2}(\tau')\hat{M}_{3';\p_3}(\tau') \delta^3(\p_1+\p_2+\p_3)\ ,
\end{equation}
where, $\lambda(\tau')$ and $f(\p_1,\p_2,\p_3)$ are functions that we will keep unspecified. $ \hat{M}_{n;\p_n}(\tau')$'s are either scalar and tensor perturbations $\hat{\R}_{\k}(\tau')$ and $\hat{h}^s_{\k}(\tau')$ or  their derivatives $\partial_{\tau'}\hat{\R}_{\k}(\tau')$ and $\partial_{\tau'}\hat{h}^{s}_{\k}(\tau')$ (in the interaction picture). Similar to (\ref{free}), they can be expressed in the following way
\begin{equation}
\hat{M}_{n;\k_n}(\tau')=\frac{1}{\sqrt{2}}\left[{\hat{c}^s}_{\k_n}v^*_{n; k_n}(\tau')+\hat{c}^{s\dagger}_{-\k_n}v_{n;k_n}(\tau')\right]+\bar{M}_{n;\k_n}(\tau')\ ,
\end{equation}
where $v_{n;k_n}(\tau')$ is the mode function associated with the operator $\hat{M}_{n;\k_n}(\tau')$ and $\bar{M}_{n;\k_n}(\tau')=\langle C|\hat{M}_{n;\k_n}(\tau')|C\rangle$.

 Now, let us evaluate the quantity (in the leading order)\footnote{Note that $\lambda(\tau')$ in $H_{int}(\tau')$ is already slow-roll suppressed and hence we only need the leading contribution.}
 \begin{align}
 \langle C|&\left[\hat{O}^I_{1;\k_1}(\tau)\hat{O}^I_{2;\k_2}(\tau)\hat{O}^I_{3;\k_3}(\tau), \hat{M}_{1';\p_1}(\tau')\hat{M}_{2';\p_2}(\tau')\hat{M}_{3';\p_3}(\tau')\right]|C\rangle\nonumber\\
&~~~~~~~~~~~~~~ =\langle 0|\left[\hat{O}^I_{1;\k_1}(\tau)\hat{O}^I_{2;\k_2}(\tau)\hat{O}^I_{3;\k_3}(\tau), \hat{M}_{1';\p_1}(\tau')\hat{M}_{2';\p_2}(\tau')\hat{M}_{3';\p_3}(\tau')\right]|0\rangle \nonumber\\
 &~~~~~~~~~~~~~~~+ \left(\langle 0|\hat{O}^I_{1;\k_1}(\tau)\hat{O}^I_{2;\k_2}|0\rangle\langle 0|\left[\hat{O}^I_{3;\k_3}(\tau), \hat{M}_{1';\p_1}(\tau')\right]|0\rangle \bar{M}_{2';\p_2}(\tau')\bar{M}_{3';\p_3}(\tau')\right.\nonumber\\
 &~~~~~~~~~~~~~~~~~~~~~~~~~~~~~~~~~~~~~~~~~~~~~~~+\text{cyclic perm} (1',2',3')+\text{cyclic perm} (1,2,3)\Big)\ .
 \end{align} 
 One can also check that in the first order in perturbation theory
 \begin{align}
 \langle C|\hat{O}_{1;\k_1}(\tau)|C\rangle=&-i \int_{\tau_0}^{\tau}d\tau'\lambda(\tau') \int d^3\p_1 d^3\p_2 d^3\p_3 f(\p_1,\p_2,\p_3)\delta^3(\p_1+\p_2+\p_3)\\
& \times \left(\langle 0|\left[\hat{O}^I_{1;\k_1}(\tau), \hat{M}_{1';\p_1}(\tau')\right]|0\rangle \bar{M}_{2';\p_2}(\tau')\bar{M}_{3';\p_3}(\tau')+\text{cyclic perm} (1',2',3')\right)\ . \nonumber
\end{align} 
 Finally one can easily show that
 \begin{align}
\langle C|\hat{O}_{1;\k_1}(\tau)\hat{O}_{2;\k_2}(\tau)&\hat{O}_{3;\k_3}(\tau)|C\rangle=\langle 0|\hat{O}_{1;\k_1}(\tau)\hat{O}_{2;\k_2}(\tau)\hat{O}_{3;\k_3}(\tau)|0\rangle_{phy}\nonumber\\
&+\left(\langle C|\hat{O}_{1;\k_1}(\tau)|C\rangle \langle C|\hat{O}_{2;\k_2}(\tau)\hat{O}_{3;\k_3}(\tau)|C\rangle+ \text{cyclic perm}(1,2,3)\right)
\end{align}
Therefore, in the tree-level, using equation (\ref{effthreepointfun}) we obtain
 \begin{equation}
\langle C|\hat{O}_{1;\k_1}(\tau)\hat{O}_{2;\k_2}(\tau)\hat{O}_{3;\k_3}(\tau)|C\rangle_{phy}=\langle 0|\hat{O}_{1;\k_1}(\tau)\hat{O}_{2;\k_2}(\tau)\hat{O}_{3;\k_3}(\tau)|0\rangle_{phy}\ .
\end{equation}
Therefore, the non-Gaussianity matrix $\F$ remains the same\footnote{One can check that the power-spectrums with coherent states are identical to that with the Bunch-Davies state and hence this is consistent with equation (\ref{F}).}
\begin{equation}
\F\equiv
 \begin{pmatrix}
  \F_{\R\R}&~~~~~  \F_{\R h}  \\
  \F_{h\R} &~~~~ \F_{hh} 
 \end{pmatrix}
 =
  \begin{pmatrix}
  -6\epsilon+2\eta ~~~& 3/2  \\
  -2\epsilon ~~& 3/2
 \end{pmatrix}\ .
\end{equation}

 \section{Non-Gaussianities from $\alpha$-states: violation of consistency relations}\label{alpha}
 In this section, we will compute the three-point functions with another special class of excited states $|\alpha\rangle$; these states are related to the Bunch-Davies state by Bogoliubov transformations and we will call them $\alpha$-states.\footnote{These states are also called Bogoliubov states.} These states are annihilated by operator $\hat{b}^s_\k$:
 \begin{equation}
 \hat{b}^s_\k|\alpha\rangle=0\ , \qquad \text{where} \qquad \hat{b}^s_\k=\alpha_s^*(k)\hat{a}^s_\k+\beta_s(k){\hat{a}_{-\k}}^{s\dagger}
 \end{equation}
 for $s=0,\times,+$. $\alpha_s(k)$ and $\beta_s(k)$ are arbitrary complex functions of $k$ (for simplicity we consider them to be function of the magnitude only) that satisfy
 \begin{equation}
 |\alpha_s(k)|^2-|\beta_s(k)|^2=1\qquad \text{for} \qquad s=0,\times,+\ .
 \end{equation}
 A state $|\alpha\rangle$ can be written explicitly as an excited state built over the Bunch-Davies state in the following way
 \begin{equation}
 |\alpha\rangle=\left[\prod_{s=0,\times,+}\prod_{\k}\frac{1}{|\alpha_s(k)|^{1/2}}\exp\left(-\frac{\beta_s(k)}{2\alpha_s^*(k)}{\hat{a}_{\k}}^{s\dagger}{\hat{a}_{-\k}}^{s\dagger}\right)\right]|0\rangle\ .
 \end{equation}
Few comments are in order: it can be shown that $\alpha$-states are normalizable only if $|\beta_s(k)|^2\rightarrow 0$ faster than $k^{-3}$ at $k\rightarrow \infty$. However, the condition that these states do not introduce any new divergences to the energy-momentum tensor requires  $|\beta_s(k)|^2\rightarrow 0$ faster than $k^{-4}$ for large $k$ (see section \ref{genstate}).

In these states, it is more convenient to express scalar and tensor perturbations in terms of operators  $\hat{b}^s_\k$ and $\hat{b}^{s\dagger}_\k$:
\begin{equation}
\hat{\R}_{\k}(\tau)=\frac{1}{\sqrt{2}}\left[\hat{b}^0_{\k}\tilde{\R}^*_k(\tau)+ \hat{b}^{0\dagger}_{-\k}\tilde{\R}_k(\tau)\right], \qquad \hat{h}^s_{\k}(\tau)=\frac{1}{\sqrt{2}}\left[\hat{b}^s_{\k}\tilde{h}^{s*}_k(\tau)+ \hat{b}^{s\dagger}_{-\k}\tilde{h}^s_k(\tau)\right]
\end{equation}
where,
\begin{align}
\tilde{\R}_k(\tau)=&  \left(\frac{H}{a \dot{\bar{\phi}}} \right)\left(\alpha_0 (k)u_k(\tau)+\beta_0 (k)u^*_k(\tau)\right),\quad \tilde{h}^s_k(\tau)=\left(\frac{\sqrt{2}}{a \M} \right)\left(\alpha_s (k)u_k(\tau)+\beta_s (k)u^*_k(\tau)\right),
\end{align}
with $u_k(\tau)=\frac{e^{i k \tau}}{\sqrt{k}}\left(1+\frac{i}{k\tau}\right)$. Practically, computations with $\alpha$-states are similar to that with the Bunch-Davies state but we have to replace the mode function $u_k(\tau)$ by $\alpha_s (k)u_k(\tau)+\beta_s (k)u^*_k(\tau)$ with appropriate $s$.

In this section, we will keep the discussion general and not specify the functional forms of $\beta_s(k)$. It is important to note that in this section we will assume that the energies of these states are not large enough to affect the slow-roll parameters. 
The power spectrum and the spectral index of scalar perturbations with $\alpha$-states are obtained to be 
\begin{eqnarray}
P_\R(k) =  \frac{H^4}{2\dot{\phi}^2k^3} \left| \alpha_0(k) - \beta_0(k) \right|^2\ , \quad n_s - 1 = 2 \eta - 6 \epsilon + \frac{d}{d \ln k}\ln \left|\alpha_0(k) - \beta_0(k) \right|^2.
\end{eqnarray}
Similarly, the power spectrum and the spectral index of tensor perturbations with $\alpha$-states are obtained to be 
\begin{eqnarray}
P_h (k) = \frac{1}{k^3} \frac{H^2}{M_{pl}^2} \left| \alpha_s(k) - \beta_s(k) \right|^2, \quad n_t= - 2 \epsilon + \frac{d}{d \ln k}\ln\sum_{s=+,\times} \left|\alpha_s(k) - \beta_s(k) \right|^2.
\end{eqnarray}
Next we will calculate the three-point functions in the squeezed limit with $\alpha$ states to show that they still can be written as (\ref{cr}), however, consistency relation (\ref{F}) is  violated. In this section we will only present the squeezed limit results; general results are relegated to appendix \ref{appendix}.

\subsection{Three scalars correlator}
The calculation for the scalar three-point function with $\alpha$-state as the initial state is identical to the  computation of section (\ref{bd3pt}) and hence we only present the result. The interaction Hamiltonian has already been computed (\ref{Hint});  the redefined field $\R_c$ is given by equation (\ref{rc}). In the squeezed limit ($k_3<<k_1=k_2$), we obtain
\begin{align}
\langle \alpha|\hat{\R}_{\k_1}(\tau)\hat{\R}_{\k_2}&(\tau)\hat{\R}_{\k_3}(\tau)|\alpha\rangle_{phy}\approx(2 \pi)^3 P_R(k_3)P_R(k_1)\left[4\epsilon\left(\frac{k_1}{k_3}\right)\Phi(k_1,k_3)-6\epsilon+2\eta\right]\delta^3(\sum{\k}).
\end{align}
Therefore, in the squeezed limit, $f_{NL}^{loc}$ is given by,
\begin{equation}\label{fnl}
f_{NL}^{loc}\approx \frac{5}{12}\left[-4\epsilon\left(\frac{k_1}{k_3}\right)\Phi(k_1,k_3)+6\epsilon-2\eta\right]\ ,
\end{equation}
where $\Phi(k_1,k_3)$ is given by,
\begin{equation}
\Phi(k_1,k_3)=\alpha_0(k_1)\beta_0(k_1)\left(\frac{\alpha^*_0(k_1)-\beta^*_0(k_1)}{\alpha_0(k_1)-\beta_0(k_1)}\right)\left(\frac{\alpha_0(k_3)+\beta_0(k_3)}{\alpha_0(k_3)-\beta_0(k_3)}\right)+c.c.
\end{equation}
In general the first term in equation (\ref{fnl}) is large in the limit $k_3<< k_1$ and hence the consistency condition is violated. In section \ref{brn}, we will estimate how large this violation can be. But before that let us comment on why the consistency relation is violated. For the derivation of the consistency relations, it is necessary to take the squeezed limit first and then calculate the three-point functions. However, in an honest calculation of the squeezed limit three-point function for a particular model, one should compute the three-point function first and then take the squeezed limit. So, there is an implicit assumption that the terms that are ignored by taking the squeezed limit first are small. The three-point function (this is true for all the three point functions) with $\alpha$-states contains terms like (where $\tau_0$ is the conformal time in the beginning of inflation) 
\begin{equation}
 i\int_{\tau_0}^0 d\tau e^{i\tau(-k_1+k_2-k_3)}+c.c= 2\left(\frac{1-\cos (-k_1+k_2-k_3)\tau_0}{-k_1+k_2-k_3}\right)
\end{equation}
that are absent for the Bunch-Davies state. Now if we take the limit $\tau_0\rightarrow -\infty$ first and then $k_3\rightarrow 0$, we obtain
\begin{equation}
 i\int_{\tau_0}^0 d\tau e^{i\tau(-k_1+k_2-k_3)}+c.c\sim -\frac{2}{k_3}
\end{equation}
which is large in the squeezed limit. However, if incorrectly we take the limit $k_3\rightarrow 0$ first, then we obtain 
\begin{equation}
 i\int_{\tau_0}^0 d\tau e^{i\tau(-k_1+k_2-k_3)}+c.c\approx 0.
\end{equation}
Therefore, the terms that we missed by taking the squeezed limit $k_3\rightarrow 0$ first are rather large and hence the consistency relations are violated. 
\subsection{Two scalars and a graviton correlator}
The interaction Hamiltonian in the momentum space is given by equation (\ref{Hssh}). The two scalars and a graviton three-point function in the squeezed limit ($k_3<<k_1=k_2$) can be calculated easily, yielding
\begin{align}\label{ssh}
 \langle \alpha|\hat{\R}_{\k_1}(\tau)\hat{\R}_{\k_2}(\tau)\hat{h}^s_{\k_3}(\tau)|\alpha\rangle_{phy}=(2\pi)^3 \delta^3(\sum\k)P_\R(k_1) P_h(k_3) \frac{\epsilon^{s}_{ij}(\k_3)\k_{1,i}\k_{1,j}}{k_1^2}\nonumber\\
 \times\left[-2\left(\frac{k_1}{k_3}\right)\Theta(k_1,k_3)+\frac{3}{2}+...\right]\ ,
\end{align}
where, function $\Theta(k_1,k_3)$ depends on the initial state and it is given by
\begin{equation}
\Theta(k_1,k_3)=\alpha_0(k_1)\beta_0(k_1)\left(\frac{\alpha^*_0(k_1)-\beta^*_0(k_1)}{\alpha_0(k_1)-\beta_0(k_1)}\right)\left(\frac{\alpha_s(k_3)+\beta_s(k_3)}{\alpha_s(k_3)-\beta_s(k_3)}\right)+c.c.
\end{equation}
Few comments are in order. In general the first term in equation (\ref{ssh}) is large in the limit $k_3\rightarrow 0$ and hence the consistency condition is violated. However, when the scalar perturbations are initially in the Bunch-Davies state (but tensor perturbations are in an $\alpha$-state),  $\Theta(k_1,k_3)=0$ and  hence the consistency condition is respected.

Note that in the squeezed limit $k_3<<k_2=k_1$ the other three-point function 
\begin{equation}
 \frac{\langle \alpha|\hat{\R}_{\k_1}(\tau)\hat{h}^s_{\k_2}(\tau)\hat{\R}_{\k_3}(\tau)|\alpha\rangle_{phy}}{P_h(k_1) P_\R(k_3)}\approx\O\left(|\beta_s(k_1)|\frac{k_3}{k_1}\right)\ .
\end{equation}
In section \ref{brn}, we will show that $|\beta_s(k_1)|<<1$ and hence this three-point function remains vanishingly small.

\subsection{Two gravitons and a scalar correlator}
The interaction Hamiltonian in momentum space is given by equation (\ref{Hhhs}). The two scalars and a graviton three-point function in the squeezed limit ($k_3<<k_1=k_2$) can be calculated easily, yielding
\begin{align}
\langle \alpha|\hat{h}^{s}_{\k_1}(\tau)\hat{h}^{s'}_{\k_2}(\tau)\hat{\R}_{\k_3}(\tau)|\alpha\rangle_{phy}\approx  (2\pi)^3& P_h(k_1)P_{\R}(k_3)\delta_{ss'}\delta^3(\sum{\k})\ \nonumber\\
&\times\left[4\epsilon \left(\frac{k_1}{k_3}\right)\Psi(k_1,k_3)-2\epsilon+...\right]\ ,
\end{align}
where, $\Psi(k_1,k_3)$ depends on the initial state
\begin{equation}
\Psi(k_1,k_3)=\alpha_s(k_1)\beta_s(k_1)\left(\frac{\alpha^*_s(k_1)-\beta^*_s(k_1)}{\alpha_s(k_1)-\beta_s(k_1)}\right)\left(\frac{\alpha_0(k_3)+\beta_0(k_3)}{\alpha_0(k_3)-\beta_0(k_3)}\right)+c.c.
\end{equation}
Note that the consistency condition (\ref{F}) is again violated unless tensor perturbations are in the Bunch-Davies state. Whereas it is easy to check that in the squeezed limit $k_3<<k_2=k_1$, the other three-point function remains vanishingly small 
\begin{align}
\frac{\langle \alpha|\hat{h}^{s}_{\k_1}(\tau)\hat{\R}_{\k_2}(\tau)\hat{h}^{s'}_{\k_3}(\tau)|\alpha\rangle_{phy}}{P_h(k_3)P_{\R}(k_2)}\approx  \O\left( \frac{k_3}{k_1} \epsilon|\beta_0(k_1)|\right) 
\end{align}
and hence it still obeys the consistency condition (\ref{cr}).

\subsection{Three gravitons correlator}
The interaction Hamiltonian in momentum space is given by equation (\ref{Hhint}). The three gravitons three-point function in the squeezed limit ($k_3<<k_1=k_2$) can be calculated easily, yielding
\begin{align}
\langle \alpha|\hat{h}^s_{\k_1}(\tau) \hat{h}^{s'}_{\k_2}(\tau)\hat{h}^{s''}_{\k_3}(\tau)|\alpha\rangle_{phy}=(2\pi)^3 \delta^3(\sum\k)P_h(k_1) P_h(k_3) \delta_{ss'}\frac{\epsilon^{s''}_{ij}(\k_3)\k_{1,i}\k_{1,j}}{k_1^2}\nonumber\\
 \times\left[-2\left(\frac{k_1}{k_3}\right)\Theta(k_1,k_3)+\frac{3}{2}+...\right]\ .
\end{align}
Where, function $\Theta(k_1,k_3)$ depends on the initial state and it is given by
\begin{equation}
\Theta(k_1,k_3)=\alpha_s(k_1)\beta_s(k_1)\left(\frac{\alpha^*_s(k_1)-\beta^*_s(k_1)}{\alpha_s(k_1)-\beta_s(k_1)}\right)\left(\frac{\alpha_{s''}(k_3)+\beta_{s''}(k_3)}{\alpha_{s''}(k_3)-\beta_{s''}(k_3)}\right)+c.c.
\end{equation}
In general the first term in equation (\ref{ssh}) is large in the limit $k_3\rightarrow 0$ and hence the consistency condition is violated. 

Let us now consider a special case: $\beta_0(k)=\beta_+(k)=\beta_\times(k)=\beta(k)$; any pre-inflationary dynamics that excites the scalar modes will also excite the tensor modes in the same way. Therefore, the non-Gaussianity $\F$ matrix, defined in (\ref{F}), is given by,
 \begin{equation}\label{aF}
\F
  =2f(k_1,k_3)
 \begin{pmatrix}
  2\epsilon&~~~~~  -1  \\
  2\epsilon &~~~~~ -1 
 \end{pmatrix}+
  \begin{pmatrix}
  -6\epsilon+2\eta ~~~& 3/2  \\
  -2\epsilon ~~& 3/2
 \end{pmatrix}
\end{equation}
where,
\begin{align}
f(k_1,k_3)=&\left(\frac{k_1}{k_3}\right)\left[\alpha(k_1)\beta(k_1)\left(\frac{\alpha^*(k_1)-\beta^*(k_1)}{\alpha(k_1)-\beta(k_1)}\right)\left(\frac{\alpha(k_3)+\beta(k_3)}{\alpha(k_3)-\beta(k_3)}\right)+c.c\right]\ .
\end{align}
In particular $f_{NL}^{loc}$ is given by,
\begin{equation}
f_{NL}^{loc}\approx \frac{5}{12}\left[-4\epsilon f(k_1,k_3)+6\epsilon-2\eta\right]\ .
\end{equation}
Now if we want to preserve scale-invariance, the function $\beta(k)$ has to be approximately constant for all the observable modes. In that case, it is obvious that the $\F$-matrix for $\alpha$-states is not consistent with (\ref{F}) because the first term in equation (\ref{aF}) dominates in the squeezed limit $k_3<<k_1=k_2$. In the next section, we will estimate how large $f(k_1,k_3)$ can be for states with energies not too large to affect the slow-roll parameters.
\section{Constraints from back-reaction}\label{brn}
Let us now consider back-reaction of the excited initial states. Before we proceed, we have to define the energy-momentum tensor of the perturbations. The Einstein's equations for the full system is  $G_{\mu\nu}-8 \pi G T_{\mu\nu}\equiv \Pi_{\mu\nu}=0$. Following \cite{Abramo:1997hu}, we can perform a perturbative expansion of the Einstein's equations:
\begin{equation}
\Pi_{\mu\nu}=\Pi^{(0)}_{\mu\nu}+\Pi^{(1)}_{\mu\nu}+\Pi^{(2)}_{\mu\nu}+...
\end{equation}
Evolution of the background is given by the lowest order equation $\Pi^{(0)}_{\mu\nu}=0$. The first order Einstein's equations $\Pi^{(1)}_{\mu\nu}=0$ give the equations of motion for the perturbations. Therefore, we can write
\begin{equation}
G^{(0)}_{\mu\nu}=8\pi G_N T^{(0)}_{\mu\nu}-\Pi^{(2)}_{\mu\nu}+...\ ,
\end{equation}
where $\Pi^{(2)}_{\mu\nu}$ has to be computed with the perturbations that solve the equations of motion $\Pi^{(1)}_{\mu\nu}=0$. From the last equation, it is clear that the energy-momentum tensor of the perturbations is given by $8\pi G_N\T_{\mu\nu}=-\Pi^{(2)}_{\mu\nu}$. Obviously both scalar and tensor perturbations will contribute to the energy-momentum tensor:
\begin{equation}
\T_{\mu\nu}=\T^s_{\mu\nu}+\T^t_{\mu\nu}\ .
\end{equation}
Explicit forms of $\T^s_{\mu\nu}$ and $\T^t_{\mu\nu}$ for single-field inflation can be found in \cite{Abramo:1997hu}.

We will promote $\T_{\mu\nu}$ to an operator and estimate $\langle \hat{\T}_{\mu\nu} \rangle$ for single-field slow-roll inflation with $\alpha$-states. $\langle \hat{\T}_{\mu\nu} \rangle$ contains UV-divergences and hence should be properly renormalized using any regularization method (for example adiabatic regularization).\footnote{ Detailed discussions of adiabatic regularization method can be found in \cite{Parker:1974qw, Bunch:1980vc}.} For our purpose, for a general initial state $|G\rangle$ it is sufficient to define the renormalized energy momentum tensor of the fluctuations in the following way:
\begin{equation}\label{ren}
\langle G| \hat{\T}_{\mu\nu}|G\rangle_{ren}=\langle G| \hat{\T}_{\mu\nu}|G\rangle-\langle 0| \hat{\T}_{\mu\nu}|0 \rangle\ 
\end{equation}
since a well-behaved initial state should not introduce any new ultra-violet divergences to the energy-momentum tensor. 

Our goal is not to perform an exact computation but to estimate how large $\beta_0(k)$ and $\beta_s(k)$ can be without causing large back-reaction. From that we will estimate how large the deviations from non-Gaussianity consistency relations can be for $\alpha$-states.  For a particular state, undoubtably an exact computation  will be more useful. 

Before we proceed let us explicitly write down $T^{(0)}_{\mu\nu}$ for single field inflation:
\begin{align}
T^{(0)}_{00}&=\rho^{(0)}=\frac{1}{2}\dot{\bar{\phi}}^2+V\left(\bar{\phi}\right)\ ,\\
T^{(0)}_{ij}&=\delta_{ij}a^2 p^{(0)}=\delta_{ij}a^2 \left(\frac{1}{2}\dot{\bar{\phi}}^2-V\left(\bar{\phi}\right)\right)\ .
\end{align}
In the beginning of inflation i.e. at $\tau=\tau_0$, following \cite{Kundu:2011sg} the leading contribution to the energy-momentum tensor of scalar fluctuations is given by 
\begin{align}
\langle \hat{\T}^s_{00}\rangle \approx&\frac{1}{2}\left(\frac{\dot{\bar{\phi}}}{Ha}\right)^2\left[\langle(\hat{\R}')^2\rangle+\langle(\nabla \hat{\R})^2\rangle\right]\ ,\\
\langle \hat{\T}^s_{ij}\rangle \approx&  \delta_{ij}\left(\frac{\dot{\bar{\phi}}}{H}\right)^2\left[\frac{1}{2}\langle(\hat{\R}')^2\rangle- \frac{1}{6}\langle(\nabla \hat{\R})^2\rangle\right]\ .
\end{align}
Similarly, the leading contribution to the energy-momentum tensor of  tensor fluctuations is given by (see section \ref{genstate})
\begin{align}
\langle \hat{\T}^t_{00}\rangle\approx&\frac{\M^2}{8 a^2}\left[\langle({\hat{h}'}_{kl})^2\rangle+\langle(\partial_m {\hat{h}}_{kl})^2 \rangle\right]\ ,\\
\langle \hat{\T}^t_{ij}\rangle=& \frac{3\M^2}{8} \delta_{ij}\left[-\langle({\hat{h}'}_{kl})^2\rangle+\langle(\partial_m {\hat{h}}_{kl})^2 \rangle\right]\nonumber\\
&~~~+\M^2\left[\frac{1}{2} \langle{\hat{h}'}_{ik}{\hat{h}'}_{kj}\rangle+\frac{1}{4}\langle(\partial_i {\hat{h}}_{kl})(\partial_j {\hat{h}}_{kl})\rangle-\frac{1}{2}\langle(\partial_l {\hat{h}}_{ki})(\partial_l {\hat{h}}_{jk})\rangle\right].
\end{align}
We will now compute these quantities for $\alpha$-states. We will assume that both $\beta_0(k)$ and $\beta_s(k)$ are nonzero and approximately constant for $k_0<k<k_*$, where, $k_0=a_0 H$, $a_0$ being the scale factor at the initial time $\tau=\tau_0$. For $k>k_*$,  $\beta_0(k)$ and $\beta_s(k)$ drop to zero very fast.\footnote{States like these are relevant if we want to preserve the scale invariance of scalar and tensor power spectrums.} We have assumed that modes inside the horizon ($k>k_0$) at $\tau=\tau_0$ are uncorrelated with modes outside the horizon ($k<k_0$) and only modes inside the horizon are excited at $\tau=\tau_0$ by some pre-inflationary causal dynamics.  For $k_0<k<k_*$, spectral indices remain unchanged 
\begin{equation}
n_s\approx 1-6\epsilon+2\eta \ , \qquad n_t\approx -2 \epsilon\ .
\end{equation}
Note that the squeezed limit three-point functions will have the nontrivial $k_1/k_3$ term only when $k_1<k_*$. Let us now compute the renormalized energy-momentum tensor of scalar  fluctuations at $\tau=\tau_0$.
\begin{align}
\langle\alpha| \hat{\T}^s_{00}|\alpha\rangle \approx \frac{1}{4\pi^2 a_0^4}\int^{k_*}k^3 dk \left(1+2|\beta_0(k)|^2\right)\ ,
\end{align}
where again $a_0$ is the scale factor at the initial time $\tau=\tau_0$. Note that we have ignored the terms with exponential factors $e^{2ik\tau_0}$ or $e^{-2ik\tau_0}$ because they oscillate rapidly. Now using (\ref{ren}), we obtain,
\begin{equation}
\langle\alpha| \hat{\T}^s_{00}|\alpha\rangle_{ren} \approx \frac{1}{2\pi^2 a_0^4}\int^{k_*}k^3 dk |\beta_0(k)|^2\approx \frac{H^4}{8\pi^2 }\left(\frac{k_*}{k_0}\right)^4|\beta_0(k_*)|^2\ ,
\end{equation} 
where $k_0=a_0H$. Similarly for other components of the energy-momentum tensor, we obtain
\begin{equation}
\langle\alpha| \hat{\T}^s_{ij}|\alpha\rangle_{ren} \approx \delta_{ij}\frac{a_0^2 H^4}{24\pi^2 }\left(\frac{k_*}{k_0}\right)^4|\beta_0(k_*)|^2\ .
\end{equation}
We can perform a similar computation for tensor perturbations and at $\tau=\tau_0$ we obtain
\begin{align}
\langle\alpha| \hat{\T}^t_{00}|\alpha\rangle_{ren}&\approx \frac{H^4}{8\pi^2 }\left(\frac{k_*}{k_0}\right)^4\sum_{s=+,\times}|\beta_s(k_*)|^2\ ,\\
\langle\alpha| \hat{\T}^t_{ij}|\alpha\rangle_{ren} &\approx \delta_{ij}\frac{a_0^2 H^4}{24\pi^2 }\left(\frac{k_*}{k_0}\right)^4\sum_{s=+,\times}|\beta_s(k_*)|^2\ .
\end{align} 
Before we proceed few comments are in order. Note that both scalar and tensor perturbations behave like radiation and their energy densities decay as $1/a^4$. And also one can check that 
\begin{equation}
p^s=\frac{1}{3}\rho^s\ , \qquad p^t=\frac{1}{3} \rho^t\ ,
\end{equation}
as expected for radiations. When initial states of scalar and tensor perturbations are the same i.e. $\beta_s(k)=\beta_0(k)$, it is easy to show that
\begin{equation}
\rho^t=2\rho^s\ , \qquad p^t=2 p^s\ ,
\end{equation}
hence tensor perturbations contribute more to the energy-momentum tensor. 

The back-reaction will not alter the background evolution if $T^{(0)}_{\mu\nu}>>\langle \hat{\T}_{\mu\nu} \rangle$. For slow-roll inflation $\frac{1}{2}\dot{\bar{\phi}}^2<<V\left(\bar{\phi}\right)$ and hence the energy densities $\rho^s$ and $\rho^t$ must be small compare to the kinetic energy of inflation for the background evolution to remain unaltered \cite{Flauger:2013hra}.\footnote{It is impotent to note that as long as $\rho^s+\rho^t<<V\left(\bar{\phi}\right)$, we will have slow-roll inflation. However, the slow-roll parameter
\begin{equation}
\epsilon=-\frac{\dot{H}}{H^2}=\frac{\dot{\bar{\phi}}^2}{2H^2 \M^2}+\frac{H^2}{6\pi^2 \M^2}\left(\frac{k_*}{k_0}\right)^4\left(\frac{a_0}{a}\right)^4\sum_{s=0,+,\times}|\beta_s(k_*)|^2
\end{equation}
is now affected by the excited state when the second term is comparable to the first term. It will not affect the background evolution but it will influence the evolution perturbations and hence it should be treated more carefully. } That leads to 
\begin{equation}\label{br}
\sum_{s=0,+,\times}|\beta_s(k_*)|^2<<\frac{4\pi^2\dot{\bar{\phi}}^2}{H^4}\left(\frac{k_0}{k_*}\right)^4\ .
\end{equation}
For simplicity, we will assume  that $\beta_s(k)=\beta_0(k)=\beta(k)$, which is reasonable because any pre-inflationary dynamics that excites the scalar modes will also excite the tensor modes in the same way. Therefore from the last equation we obtain,
\begin{equation}
|\beta(k_*)|^2<<\frac{4\pi^2\dot{\bar{\phi}}^2}{3H^4}\left(\frac{k_0}{k_*}\right)^4\ .
\end{equation}
Using the fact\footnote{Recall that our squeezed limit corresponds to $k_3<<k_1=k_2$.} that $k_1<k_*$ and $k_3>k_0$ and 
\begin{equation}
\Delta_\R^2=\frac{H^4|\alpha(k_*)-\beta(k_*)|^2}{4\pi^2\dot{\bar{\phi}}^2}
\end{equation}
where, we also have the usual condition $|\alpha(k_*)|^2-|\beta(k_*)|^2=1$, we obtain
\begin{equation}
|\beta(k_*)|^2<<\frac{|\alpha(k_*)-\beta(k_*)|^2}{3\Delta_\R^2}\left(\frac{k_3}{k_1}\right)^4\ .
\end{equation}
As explained in \cite{Flauger:2013hra}, $|\beta(k_*)|>>1$ has already been ruled out if we want to avoid a step in the scalar power spectrum. For $|\beta(k_*)|<<1$ or $|\beta(k_*)|\sim\O(1)$ it is easy to check that $|\alpha(k_*)-\beta(k_*)|^2\sim \O(1)$ and we obtain
\begin{equation}
|\beta(k_*)|<<\frac{1}{\sqrt{3}\Delta_\R}\left(\frac{k_3}{k_1}\right)^2\ .
\end{equation}
Note that from observation $\Delta_\R^2=2.2\times 10^{-9}$ \cite{Ade:2013uln} and even with $k_3/k_1\sim 10^{-2}$, from the last equation we get $|\beta(k_*)|<<1$. Therefore with this constraints, we obtain
\begin{align}
f(k_1,k_3)&=\left(\frac{k_1}{k_3}\right)\left[\alpha(k_1)\beta(k_1)\left(\frac{\alpha^*(k_1)-\beta^*(k_1)}{\alpha(k_1)-\beta(k_1)}\right)\left(\frac{\alpha(k_3)+\beta(k_3)}{\alpha(k_3)-\beta(k_3)}\right)+c.c\right]\nonumber\\
&<< 2\left(\frac{k_3}{k_1}\right)\frac{1}{\sqrt{3}\Delta_\R}\ .
\end{align}
Note that the last equation is linear in $k_3/k_1$ and hence $f(k_1,k_3)\rightarrow 0$ in the limit $k_3/k_1\rightarrow 0$. However, the consistency relations are violated for the physically relevant case, i.e.  when $k_3/k_1$ is small but finite; using the observed value of $\Delta_\R^2$ and $k_3/k_1\sim 10^{-2}$, we finally get,
\begin{equation}
f(k_1,k_3)<<200\ .
\end{equation}
For three-scalars correlator, this corresponds to $f_{NL}^{loc}<<1$ and hence it is unobservable in the near future \cite{Flauger:2013hra}.\footnote{Even for $f(k_1,k_3)\sim 200$, the signal to noise ratio for Planck is $S/N <0.9$ and hence can not be detected.} However, $f(k_1,k_3)$ is large enough to violate all the consistency relations.
\section{Conclusions}\label{conclusions}
In this paper, we have studied the consistency relations of the two-point and the three-point functions of scalar and tensor perturbations in single-field inflation with general initial conditions for the perturbations. The three-point functions of the perturbations, in the squeezed limit (i.e. $k_1, k_2\gg k_3$), are known to obey certain consistency relations which are of the form
\begin{equation}
\langle \hat{A}_{\k_1}\hat{B}_{\k_2}\hat{C}_{\k_3}\rangle=(2\pi)^3 \F_{AC}P_A(k_1)P_C(k_3)\delta_{s(A),s(B)}\frac{\epsilon^{s(C)}_{ij}(\k_3)k_{1;i}k_{i;j}}{k_1^2}\delta^3\left(\sum\k\right)\ ,\nonumber
\end{equation}
where $\hat{A}_{\k},\hat{B}_{\k}, \hat{C}_{\k}$ are either scalar perturbation $\hat{\R}_{\k}$ or tensor perturbation $\hat{h}^s_{\k}$. We have used the notation that for the scalar perturbations $s(\R)=0$ and $\epsilon^{0}_{ij}(\k)\equiv\delta_{ij}$. For the tensor perturbations, $s(h)$ is the polarization of the mode and $\epsilon^s_{ij}(\k)$ is the polarization tensor. $\F_{AC}$ is a measure of non-Gaussianity and it is given by
\begin{equation}\nonumber
\F\equiv
 \begin{pmatrix}
  \F_{\R\R}&~~~~~  \F_{\R h}  \\
  \F_{h\R} &~~~~ \F_{hh} 
 \end{pmatrix}
 =
  \begin{pmatrix}
  n_s-1 ~~~& 2-\frac{n_s}{2}  \\
  n_t & \frac{3-n_t}{2}
 \end{pmatrix}\ .
\end{equation}
For slow-roll inflation, we find that all the three-point functions of scalar and tensor perturbations with a coherent state as the initial state are identical to three-point functions with the Bunch-Davies initial state. On the other hand, there is a violation of the consistency relations for $\alpha$-states, which are states that are related to the Bunch-Davies state by Bogoliubov transformations. 

For slow-roll inflation, tensor tilt and tensor-to-scalar- ratio obey certain consistency relation: $ r+8n_t=0$. When the perturbations are initially in excited states generated by  some pre-inflationary dynamics,  the tenser-to-scalar ratio remains unchanged. However, both $n_s$ and $n_t$ get corrected and hence the consistency relation is no longer true.  It is important to note that even with excited initial states slow-roll inflation predicts   $ r+8n_t\sim\O(\epsilon,\eta)$. Detection of primordial gravitational waves will provide a crucial test for slow-roll inflation with Bunch-Davies initial condition for the perturbations.

Let us conclude by saying that although a complete characterization of the three-point functions with different initial states is a challenging task, the three-point functions for different shapes of momentum-space triangles can be a useful tool for probing the initial state. Observations made by the current generation of cosmological experiments may contain valuable information about the initial state of primordial fluctuations and that would provide a window for the physics before inflation.  
\section*{Acknowledgments}
I want to thank  Willy Fischler, Aditya Aravind, Dan Carney, Anindya Dey,  Eiichiro Komatsu, Ely D. Kovetz, Dustin Lorshbough and Joel Meyers for helpful discussions. This material is based upon work supported by the National Science Foundation under Grant Number PHY-1316033 and PHY-0969020 and by the Texas Cosmology Center, which is supported by the College of Natural Sciences and the Department of Astronomy at the University of Texas at Austin and the McDonald Observatory.

\newpage

\appendix

\section{Three-point functions with $\alpha$-states}\label{appendix}
Here we will present full expressions of all the three-point functions with $\alpha$-states.
\subsection{Three scalars correlator}
The three scalar three-point function at late time ($\tau\rightarrow 0$) is given by
\begin{align}
\langle \alpha|\hat{\R}_{\k_1}(\tau)\hat{\R}_{\k_2}(\tau)\hat{\R}_{\k_3}(\tau)|\alpha\rangle_{phy}=\left\{\frac{(2\pi)^3}{2} P_\R(k_1)P_{\R}(k_2)\left(2\eta-3\epsilon -\epsilon\frac{k_1^2+k_2^2}{k_3^2}\right)+ \text{c.p.}\right\}&\delta^3(\sum{\k})\nonumber\\ 
-\frac{(2\pi)^3H^6\delta^3(\sum{\k})}{4k_1k_2k_3\M^2\dot{\bar{\phi}}^2}\left(\sum_{i}\frac{1}{k_i^2}\right)\big[\left(\alpha_0^*(k_1)-\beta_0^*(k_1)\right)\left(\alpha_{0}^*(k_2)-\beta_{0}^*(k_2)\right)\left(\alpha_0^*(k_3)-\beta_0^*(k_3)\right)&\nonumber\\
\times\left\{ \left(\alpha_0(k_1)\alpha_{0}(k_2)\alpha_0(k_3)+\beta_0(k_1)\beta_{0}(k_2)\beta_0(k_3)\right)\frac{1}{k_1+k_2+k_3} \right.&\nonumber\\
+ \left(\alpha_0(k_1)\alpha_{0}(k_2)\beta_0(k_3)+\beta_0(k_1)\beta_0(k_2)\alpha_0(k_3)\right)\frac{1}{-k_1-k_2+k_3} &\nonumber\\
+ \left(\alpha_0(k_1)\beta_{0}(k_2)\alpha_0(k_3)+\beta_0(k_1)\alpha_{0}(k_2)\beta_0(k_3)\right)\frac{1}{-k_1+k_2-k_3} &\nonumber\\
\left. + \left(\alpha_0(k_1)\beta_{0}(k_2)\beta_0(k_3)+\beta_0(k_1)\alpha_{0}(k_2)\alpha_0(k_3)\right)\frac{1}{k_1-k_2-k_3}\right\} &\nonumber\\
+ c.c. \big]&
\end{align}
where c.p. stands for cyclic permutations and c.c. stands for complex conjugation.

\subsection{Two scalars and a graviton correlator}
The two scalars and a graviton three-point function at late time ($\tau\rightarrow 0$) is given by
\begin{align}
 \langle \alpha|\hat{\R}_{\k_1}(\tau)\hat{\R}_{\k_2}(\tau)\hat{h}^s_{\k_3}(\tau)|\alpha\rangle_{phy}=(2\pi)^3 \delta^3(\sum\k)\frac{H^6}{\M^2 \dot{\bar{\phi}}^2}\frac{ \epsilon^{s}_{ij}(\k_3)\k_{1,i}\k_{2,j}}{4(k_1k_2k_3)^3}~~~~~~~~~~~~~~~\\
 \times \left(\alpha_0^*(k_1)-\beta_0^*(k_1)\right)\left(\alpha_0^*(k_2)-\beta_0^*(k_2)\right)\left(\alpha_s^*(k_3)-\beta_s^*(k_3)\right)\nonumber\\
 \times\left\{\left(\alpha_0(k_1)\alpha_0(k_2)\alpha_s(k_3)+\beta_0(k_1)\beta_0(k_2)\beta_s(k_3)\right)\I_0(k_1,k_2,k_3)\right.~~~~~\nonumber\\
 + \left(\alpha_0(k_1)\alpha_0(k_2)\beta_s(k_3)+\beta_0(k_1)\beta_0(k_2)\alpha_s(k_3)\right)\I_1(k_1,k_2,k_3) \nonumber\\
+ \left(\alpha_0(k_1)\beta_0(k_2)\alpha_s(k_3)+\beta_0(k_1)\alpha_0(k_2)\beta_s(k_3)\right)\I_1(k_1,k_3,k_2) \nonumber \\
\left. + \left(\alpha_0(k_1)\beta_0(k_2)\beta_s(k_3)+\beta_0(k_1)\alpha_0(k_2)\alpha_s(k_3)\right)\I_1(k_3,k_2,k_1)\right\} \nonumber\\
+ c.c. \ , \nonumber 
 \end{align}
where, $c.c.$ stands for complex conjugate and 
\begin{align}
\I_0(k_1,k_2,k_3)&=\left(-k_t+\frac{k_1k_2k_3}{k_t^2}+\frac{k_1k_2+k_2k_3+k_1k_3}{k_t}\right)\ , \\
\I_1(k_1,k_2,k_3)&=\left(k_1+k_2-k_3+\frac{k_1k_2k_3}{(k_1+k_2-k_3)^2}+\frac{-k_1k_2+k_2k_3+k_1k_3}{k_1+k_2-k_3}\right)\ .
\end{align}

\subsection{Two gravitons and a scalar correlator}
Two graviton and a scalar three-point function at late time ($\tau\rightarrow 0$) is given by
\begin{align}
\langle \alpha|\hat{h}^{s}_{\k_1}(\tau)\hat{h}^{s'}_{\k_2}(\tau)\hat{\R}_{\k_3}(\tau)|\alpha\rangle_{phy}=\frac{(2\pi)^3}{16}P_h(k_1)P_{h}(k_2)\epsilon^{s}_{ij}(\k_1)\epsilon^{s'}_{ij}(\k_2)\left(\frac{k_3^2-k_1^2-k_2^2}{k_3^2}\right)\delta^3(\sum{\k})\nonumber\\ 
-\frac{(2\pi)^3H^4}{4k_1k_2k_3^3\M^4}\epsilon^{s}_{ij}(\k_1)\epsilon^{s'}_{ij}(\k_2)\delta^3(\sum{\k})\big[\left(\alpha_s^*(k_1)-\beta_s^*(k_1)\right)\left(\alpha_{s'}^*(k_2)-\beta_{s'}^*(k_2)\right)\left(\alpha_0^*(k_3)-\beta_0^*(k_3)\right)\nonumber\\
\times\left\{ \left(\alpha_s(k_1)\alpha_{s'}(k_2)\alpha_0(k_3)+\beta_s(k_1)\beta_{s'}(k_2)\beta_0(k_3)\right)\frac{1}{k_1+k_2+k_3} \right.\nonumber\\
+ \left(\alpha_s(k_1)\alpha_{s'}(k_2)\beta_0(k_3)+\beta_s(k_1)\beta_{s'}(k_2)\alpha_0(k_3)\right)\frac{1}{-k_1-k_2+k_3} \nonumber\\
+ \left(\alpha_s(k_1)\beta_{s'}(k_2)\alpha_0(k_3)+\beta_s(k_1)\alpha_{s'}(k_2)\beta_0(k_3)\right)\frac{1}{-k_1+k_2-k_3} \nonumber\\
\left. + \left(\alpha_s(k_1)\beta_{s'}(k_2)\beta_0(k_3)+\beta_s(k_1)\alpha_{s'}(k_2)\alpha_0(k_3)\right)\frac{1}{k_1-k_2-k_3}\right\} \nonumber\\
+ c.c. \big]
\end{align}

\subsection{Three gravitons correlator}
Similarly, three gravitons correlation function can be calculated in the $\alpha$-states and the final result is 
\begin{align}
 \langle \alpha|\hat{h}^s_{\k_1}(\tau) \hat{h}^{s'}_{\k_2}(\tau)\hat{h}^{s''}_{\k_3}(\tau)|\alpha\rangle_{phy}=(2\pi)^3 \delta^3(\sum\k)\frac{H^4}{2\M^4 }
\times \left(-\epsilon^{s}_{ii'}(\k_1)\epsilon^{s'}_{jj'}(\k_2)\epsilon^{s''}_{ll'}(\k_3)t_{ijl}t_{i'j'l'}\right)\nonumber\\
 \times \frac{1}{2(k_1k_2k_3)^3}\left(\alpha_{s}^*(k_1)-\beta_{s}^*(k_1)\right)\left(\alpha_{s'}^*(k_2)-\beta_{s'}^*(k_2)\right)\left(\alpha_{s''}^*(k_3)-\beta_{s''}^*(k_3)\right)\nonumber\\
 \times\left\{\left(\alpha_s(k_1)\alpha_{s'}(k_2)\alpha_{s''}(k_3)+\beta_s(k_1)\beta_{s'}(k_2)\beta_{s''}(k_3)\right)\I_0(k_1,k_2,k_3)\right.~~~~~\nonumber\\
 + \left(\alpha_s(k_1)\alpha_{s'}(k_2)\beta_{s''}(k_3)+\beta_s(k_1)\beta_{s'}(k_2)\alpha_{s''}(k_3)\right)\I_1(k_1,k_2,k_3) \nonumber\\
+ \left(\alpha_s(k_1)\beta_{s'}(k_2)\alpha_{s''}(k_3)+\beta_s(k_1)\alpha_{s'}(k_2)\beta_{s''}(k_3)\right)\I_1(k_1,k_3,k_2) \nonumber \\
\left. + \left(\alpha_s(k_1)\beta_{s'}(k_2)\beta_{s''}(k_3)+\beta_s(k_1)\alpha_{s'}(k_2)\alpha_{s''}(k_3)\right)\I_1(k_3,k_2,k_1)\right\} \nonumber\\
+ c.c. \ , 
 \end{align}
where, $c.c.$ again stands for complex conjugate and 
\begin{align}
\I_0(k_1,k_2,k_3)&=\left(-k_t+\frac{k_1k_2k_3}{k_t^2}+\frac{k_1k_2+k_2k_3+k_1k_3}{k_t}\right)\ , \\
\I_1(k_1,k_2,k_3)&=\left(k_1+k_2-k_3+\frac{k_1k_2k_3}{(k_1+k_2-k_3)^2}+\frac{-k_1k_2+k_2k_3+k_1k_3}{k_1+k_2-k_3}\right)\ .
\end{align}

\end{document}